\documentclass[english, twocolumn, 10pt, aps, superscriptaddress, floatfix, showpacs, prl, citeautoscript]{revtex4-1}

\usepackage{graphicx}
\usepackage{dcolumn}
\usepackage{bm}
\usepackage{braket}

\begin{document}

\title{Mirage Andreev spectra generated by mesoscopic leads in nanowire quantum dots}

\author{Z. Su}
\affiliation{Department of Physics and Astronomy, University of Pittsburgh, Pittsburgh PA, 15260, USA} 
\author{A. Zarassi}
\affiliation{Department of Physics and Astronomy, University of Pittsburgh, Pittsburgh PA, 15260, USA} 
\author{J.-F. Hsu}
\affiliation{Department of Physics and Astronomy, University of Pittsburgh, Pittsburgh PA, 15260, USA} 
\author{P. San-Jose}
\affiliation{Instituto de Ciencia de Materiales de Madrid (ICMM-CSIC), Cantoblanco, 28049 Madrid, Spain}
\author{E. Prada}
\affiliation{Departamento de Fisica de la Materia Condensada,
Condensed Matter Physics Center (IFIMAC) and Instituto Nicolas Cabrera, Universidad Autonoma de Madrid, E-28049 Madrid, Spain}
\author{R. Aguado}
\affiliation{Instituto de Ciencia de Materiales de Madrid (ICMM-CSIC), Cantoblanco, 28049 Madrid, Spain}
\author{E.J.H. Lee}
\affiliation{Departamento de Fisica de la Materia Condensada,
Condensed Matter Physics Center (IFIMAC) and Instituto Nicolas Cabrera, Universidad Autonoma de Madrid, E-28049 Madrid, Spain}
\author{S. Gazibegovic}
\affiliation{Eindhoven University of Technology, 5600 MB, Eindhoven, The Netherlands}
\author{R. Op het Veld}
\affiliation{Eindhoven University of Technology, 5600 MB, Eindhoven, The Netherlands}
\author{D. Car}
\affiliation{Eindhoven University of Technology, 5600 MB, Eindhoven, The Netherlands}
\author{S.R. Plissard}
\affiliation{LAAS CNRS, Universit{\'e} de Toulouse, 31031 Toulouse, France}
\author{M. Hocevar}
\affiliation{Univ. Grenoble Alpes, CNRS, Grenoble INP, Institut N\'eel, 38000 Grenoble, France}
\author{M. Pendharkar}
\affiliation{Electrical and Computer Engineering, University of California Santa Barbara, Santa Barbara, California 93106, USA}
\author{J.S. Lee}
\affiliation{California NanoSystems Institute, University of California Santa Barbara, Santa Barbara, CA 93106, USA}
\author{J.A. Logan}
\affiliation{Materials Department, University of California, Santa Barbara, Santa Barbara, CA 93106, USA}
\author{C.J. Palmstr\o m}
\affiliation{Electrical and Computer Engineering, University of California Santa Barbara, Santa Barbara, California 93106, USA}
\affiliation{California NanoSystems Institute, University of California Santa Barbara, Santa Barbara, CA 93106, USA}
\affiliation{Materials Department, University of California, Santa Barbara, Santa Barbara, CA 93106, USA}
\author{E.P.A.M. Bakkers}
\affiliation{Eindhoven University of Technology, 5600 MB, Eindhoven, The Netherlands}
\author{S.M. Frolov}
\affiliation{Department of Physics and Astronomy, University of Pittsburgh, Pittsburgh PA, 15260, USA}

\date{\today}

\begin{abstract}

We study transport mediated by Andreev bound states formed in InSb nanowire quantum dots. Two kinds of superconducting source and drain contacts are used: epitaxial Al/InSb devices exhibit a doubling of tunneling resonances, while in NbTiN/InSb devices Andreev spectra of the dot appear to be replicated multiple times at increasing source-drain bias voltages. In both devices, a mirage of a crowded spectrum is created.  To describe the observations a model is developed that combines the effects of a soft induced gap and of additional Andreev bound states both in the quantum dot and in the finite regions of the nanowire adjacent to the quantum dot. Understanding of Andreev spectroscopy is important for the correct interpretation of Majorana experiments done on the same structures.

\end{abstract}

\maketitle
The superconductor-semiconductor hybrid structures are of recent interest due to the possibility of inducing topological superconductivity accompanied by Majorana bound states (MBS) ~\cite{Beenakker:11,Alicea:RPP12,RevModPhys.87.137,Aguadoreview17}. More generally, when a semiconductor is of finite size, proximity to a superconductor gives rise to subgap quasiparticle excitations, the so-called Andreev bound states (ABS),  that appear due to successive Andreev reflections at the interfaces. Single ABS have been demonstrated in a variety of structures including self-assembled quantum dots, semiconductor nanowires, atomic break junctions, carbon nanotubes and graphene~\cite{GroveRasmussenPRL07,EichlerPRL07, ABS_Tarucha_PRL,first_ABS, ABS_Nadya, changPRL2013,bretheauNature13}. ABS exhibit many similarities to MBS, and therefore ABS can be served as a prototypical system for Majorana studies~\cite{LeeNatnano2014,LiuPRB17}. A powerful experimental method for investigating  both MBS and ABS is via tunneling, either from a nanofabricated probe or by scanning tunneling spectroscopy. The latter is typically performed on Yu-Shiba-Rusinov states which are closely analogous to ABS but originate from magnetic impurities in superconductors~\cite{Yazdani1997,RubyPRL15}.

In this paper, we focus on the mesoscopic effects within the tunneling probes. We show that the non-trivial densities of states (DOS) in the probes can drastically affect tunneling characteristics by generating multiple replicas of ABS. To experimentally investigate these effects, we use semiconductor nanowires coupled to superconductors. ABS are induced in a quantum dot by strongly coupling the dot to one superconducting contact. A second superconducting contact and a nanowire segment adjacent to it act as a tunneling probe. To explain our observations, we consider the effects of soft induced superconducting gap in the nanowire, and of additional ABS induced in nanowire segments adjacent to the dot. The surprising observation of sub-gap negative differential conductance (NDC) is found to be consistent with a peak in the density of states of the probe at zero chemical potential, which is present even at zero magnetic field. The exact origin of this anomalous density of states remains an open question. Our findings emphasize the importance of understanding the spectral structure of the measuring contacts to interpret tunneling experiments in mesoscopic systems. We expect them to be particularly relevant for the MBS search in similar nanowire devices ~\cite{MourikScience2012, DasNaturePhy2012, DengScience2016,  Chen2017, quantized_majorana,GulNatNano18}.

\begin{figure}[ht]
  \centering
  \includegraphics[width=\columnwidth]{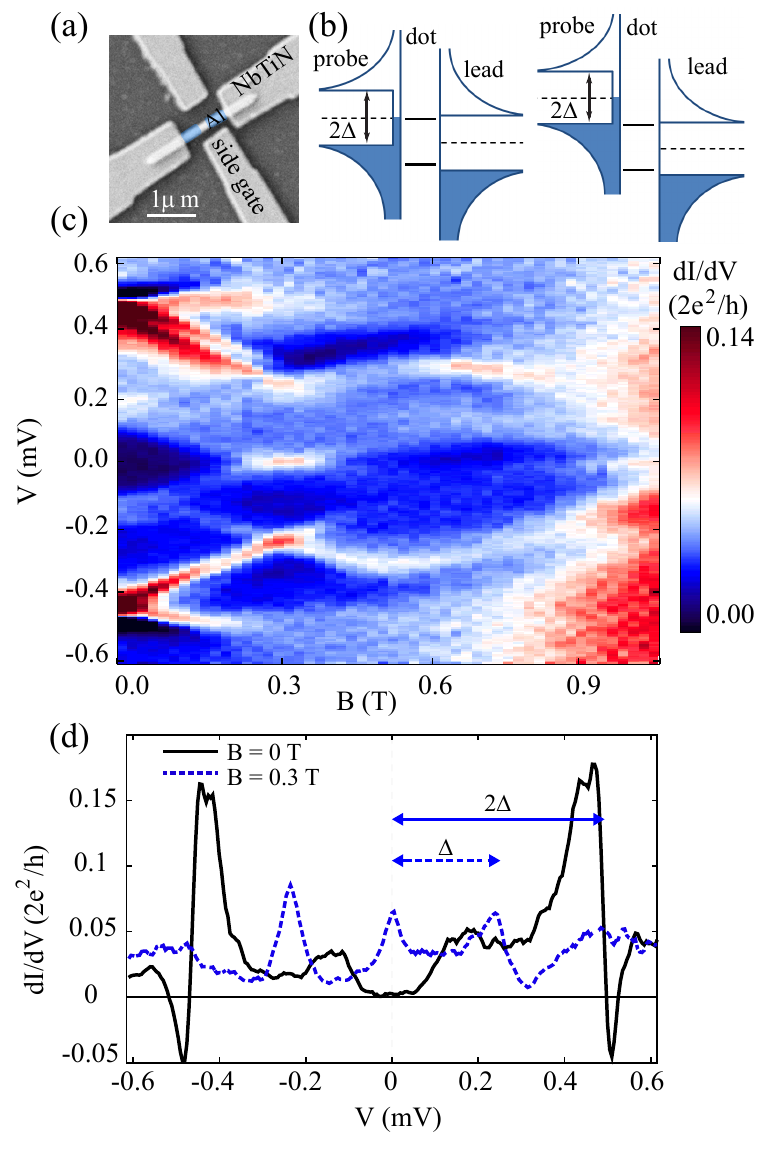}
  \caption{
(a) scanning electron micrograph of a representative Al/InSb device. The shaded blue regions show the Al thin shell with a break in the middle.  (b) illustrative energy diagrams of a soft gap probe, a hard-gapped lead and an ABS in the dot (solid lines) for two different source drain biases $V\approx\Delta/e$ (left) and $V\approx 2\Delta/e$ (right). (c) and (d) magnetic field evolution of the two-terminal transport. The field is applied parallel to the nanowire axis.}
\label{fig1}
\end{figure}

InSb nanowires are grown using metalorganic vapor phase epitaxy (MOVPE)~\cite{PlissardNanoLett12}. We investigate two devices that are drastically different both in the way they are gated and in the way superconductivity is induced. 
The first is an Al/InSb device which shows a two-replica tunneling spectrum that can be understood by only considering the effect of a soft induced gap in the nanowire. Building on the simpler example of an  Al/InSb device, we discuss the second, NbTiN/InSb, device in which multiple replicas are observed. Properly describing this effect requires a non-trivial DOS in the leads. All measurements are performed in a dilution refrigerator with a base temperature of 30 mK.

The Al/InSb device in Fig.~\ref{fig1}(a) features an epitaxially-matched thin shell of Al defined by molecular beam epitaxy (MBE), with a single break in the shell around which the quantum dot is formed~\cite{gazibegovic2017epitaxy}. The wires were allowed to age in air which possibly accounts for softer induced gap.
NbTiN contacts are fabricated on top of the Al shell of the nanowire following Ref.~\onlinecite{gazibegovic2017epitaxy}, but superconductivity in the dot is primarily induced by the Al shell since NbTiN is offset back from the break in the shell. A combination of the back and side gates is used to define a quantum dot by lowering the electron density primarily near the break in the Al shell. In practice, the side gate is fixed and only the effect of the back gate is explored (see supplemental materials for quantum dot characterization). The dot is partially defined by disorder  which becomes prominent at low density. 

In a hard-gap superconductor-superconductor tunnel junction, conductance is expected to be zero for source-drain biases $|V| < 2 \Delta/e$, where $\Delta$ is the superconducting gap which is typically $200~\mu$eV in aluminum~\cite{Pientka2015, GroveRasmussenPRB09}. 
If the probe is a soft-gap superconductor, conductance can be non-zero at lower biases. Fig.~\ref{fig1}(b) illustrates how current can flow at a bias of $V < \Delta/e$ if a small DOS is present in the probe within the superconducting gap. Another current peak is expected when the gap edge of the probe is aligned with the ABS in the dot, therefore the same ABS is responsible for two peaks in transport. 

In the Al/InSb device the conductance is non-zero for $|V| \gtrsim \Delta/e$, and two small conductance peaks are found at $V \approx \pm \Delta/e$ (Figs.~\ref{fig1}(c),(d)) at zero applied magnetic field. We argue that conductance in the range $\Delta/e < V < 2 \Delta/e$ is due to the soft gap effect which makes tunneling possible when the center of the induced gap in the probe is aligned with ABS level in the dot located close to the gap edge, as in Fig.~\ref{fig1}(b). Still, the largest peaks at zero field are at $\pm 2 \Delta$, which indicates that the subgap density of states is relatively small. The resonances at $\pm 2 \Delta$ are accompanied by negative differential conductance (NDC) shadows around $\sim \pm 0.5$ mV, which is typical for tunneling transport between two superconducting gap edges and arises due to a convolution of two DOS peaks  \cite{LeePRL2012}.

The conductance peaks at $\pm\Delta$ and $\pm 2 \Delta$ evolve in magnetic field.
Both resonances split into two branches, one of which moves to higher bias, while the other moves to lower bias. This indicates that we are observing Zeeman splitting of an ABS that is localized near the gap edge at zero field \cite{LeePRL2012}. The spectrum is doubled because the same ABS is probed by the large density of states in the probe at $V = \Delta/e$ and by the small density of states at $V = 0$. This is confirmed by that fact that the branches originating from $\Delta$ are parallel to branches originating at $2 \Delta$ at low field. Thus we are observing two replicas of the same Andreev spectrum evolving in magnetic field, offset in bias by $V = \Delta/e$. 

At B = 0.3 T resonances that originated from $\pm \Delta$ coalesce at zero bias, resulting in a zero-bias peak \cite{LeeNatnano2014}. At the same field kinks are observed in higher bias resonances around $V = \Delta/e$. The kinks appear because the positive and negative bias segments are shifted to $+\Delta$ and $-\Delta$ respectively by the probe at the gap edge. The superconducting gap in the Al shell remains virtually unchanged at B = 0.3 T. This is evidenced by the fact that the upper branch of the 1$\Delta$ resonance meets exactly with the lower branch of the $2\Delta$ resonance at that field. However, the gap collapses at higher fields and vanishes at $B \approx 1.0$ T. Therefore the two replicas look less similar at higher fields. We also note that the upper branch at $+2\Delta$ appears to split into three resonances at small fields, with two of the branches moving down, a non-universal effect which remains to be understood.

\begin{figure}
  \centering
  \includegraphics[width=\columnwidth]{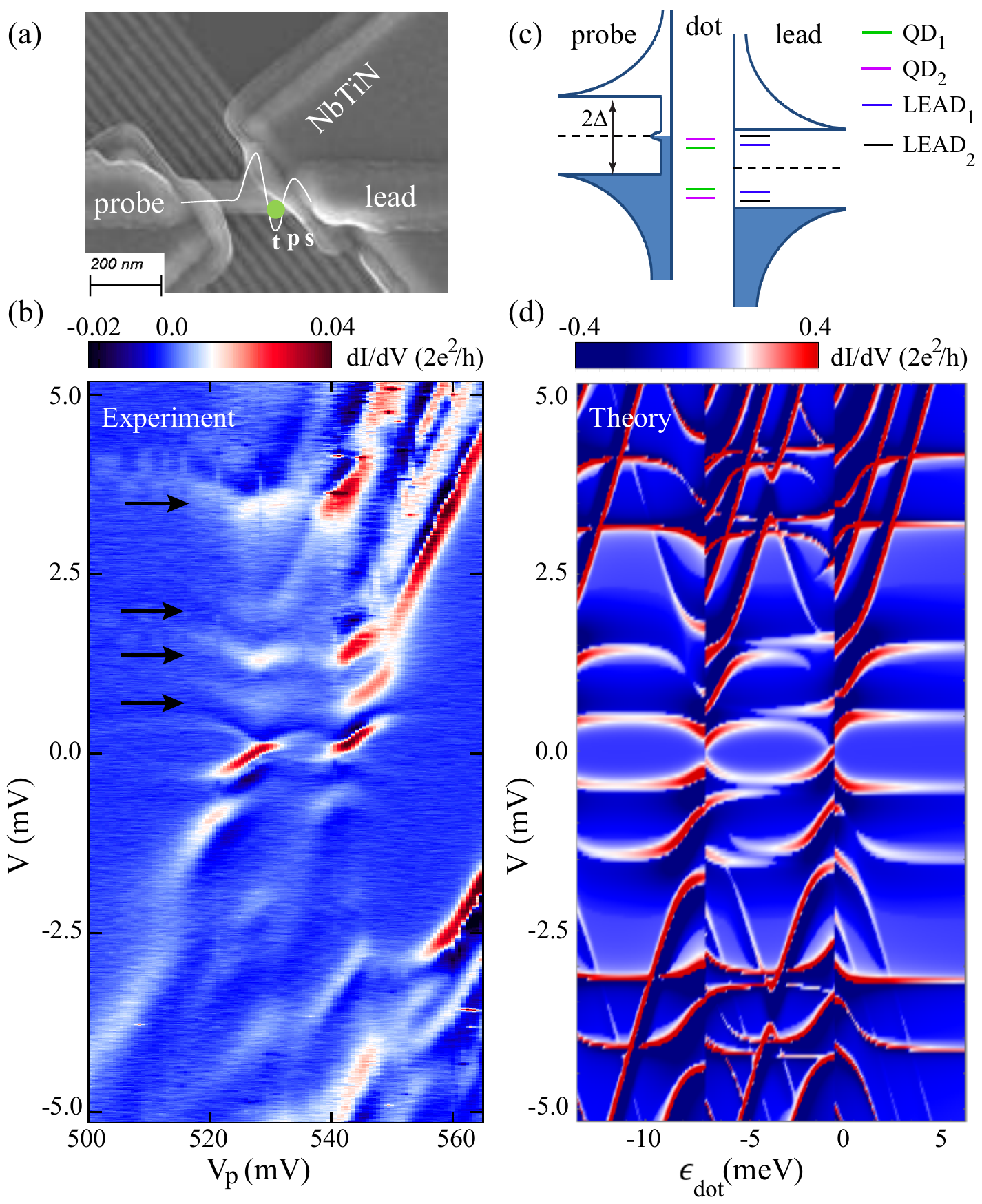}
  \caption{
  (a) Scanning electron micrograph of the NbTiN/InSb device. Green dot marks the quantum dot, white line is a conceptual confining potential set by gates $t$, $p$ and $s$. (b) Tunneling conductance through the dot as a function of bias and $V_p$. Arrows point to four apparent replicas of the lowest loop-like resonance. (c) Illustrative energy diagram with the soft gap probe, two ABS on the dot (QD$_1$ and QD$_2$) and two ABS in the hard gap lead (LEAD$_1$ and LEAD$_2$). 
(d) Theoretical model results as a function of dot on-site energy $\epsilon_\mathrm{dot}$, with QD$_{1,2}$ energies $\epsilon^D_1=\epsilon_\mathrm{dot}$ and $\epsilon^D_2=\epsilon_\mathrm{dot}-1.7$ meV, LEAD$_{1,2}$ energies $\epsilon^L_1=0.5$ meV and $\epsilon^L_2=1.5$ meV, induced pairing $\Gamma_S=0.27$ meV, parent gap $\Delta_p=2.7$ meV and Coulomb energy $U=6.8$ meV (see supplemental materials for model details).
}
 \label{fig2}
\end{figure}

Having understood the doubling of tunneling resonances due to the soft gap effect, we now discuss the less trivial behavior of the NbTiN/InSb device in which more than two apparent replicas are observed (Fig.~\ref{fig2}(a)). In this device no epitaxial Al shell is present and the nanowire directly contacts the NbTiN electrodes. 
This device is fabricated atop of an array of fine local electrostatic gate electrodes with the center-to-center distance of 60 nm.  The gate dielectric is a 10 nm thick layer of HfO$_2$. The quantum dot is fully defined by local gates labeled $t$, $p$ and $s$ for ``tunneling", ``plunger" and ``superconductor". The dot is defined close to the right superconductor and the barrier above gate $s$ is tuned so as to strongly couple the right superconductor and the dot. The left superconductor is separated from the dot by a segment of a nanowire and a high tunneling barrier defined above gate $t$. We vary the occupation of the dot with voltage $V_p$ on the plunger gate. This device has been used in a previous study for which two dots were defined in the same nanowire~\cite{su_andreevmolecule}.

Data in Fig.\ref{fig2}(b) show transport through the NbTiN/InSb device as a function of plunger gate up to a high bias of 5 mV. The lowest bias resonances (closest to zero) exhibit behavior typical for ABS in quantum dots: they form a ``loop" by crossing zero bias twice at approximately $V_p = 520$ mV and $V_p = 540$ mV. 
This is explained by the dot undergoing a singlet-doublet ground state transition at the nodes of the loop \cite{first_ABS,ABS_Tarucha_PRL,changPRL2013,LeeNatnano2014}. Interestingly, four apparent resonances that follow the same behavior of the upper half-loop are observed at increasing values of positive bias in the gate range. The highest bias resonance is at an energy consistent with twice the gap of bulk NbTiN, which has been measured to be close to 2 meV (data not shown).
Multiple Andreev reflections are known to generate a series of subgap features, but this effect is typically observed in symmetric structures, while here $p$ and $t$ barriers are tuned to be highly asymmetric. We also notice that the loop-like resonances at the center of the gate range evolve smoothly into diagonal lines, most clearly for $V_p = 540-560$ mV. These diagonals resemble excited states of a quantum dot. This is not expected for multiple Andreev reflection.

\begin{figure} 
  \centering
  \includegraphics[width=0.5\textwidth]{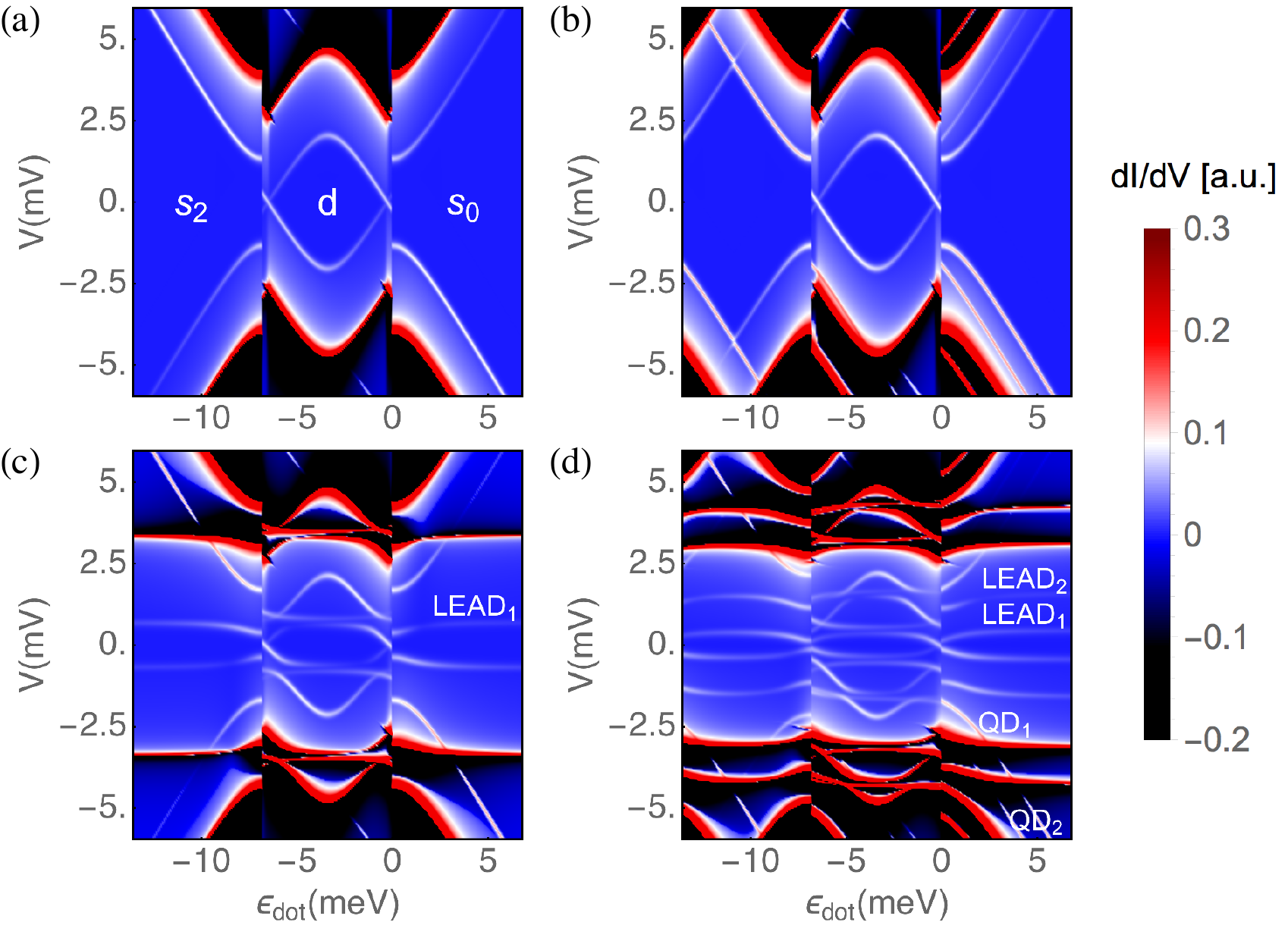}
\caption{Tunneling differential conductance at zero field across a quantum dot between a soft-gap superconducting electrode and a proximitized nanowire lead with a hard gap. The quantum dot has one (a) or two (b-d) spinful levels, while the nanowire has zero (a,b), one (c) or two (d) subgap Andreev bound states. Magnetic field is zero in all panels, simulation parameters similar to those in Fig.~\ref{fig2}(d). See Fig.~\ref{figsupp8} in the supplemental materials for details on the corresponding energy spectra.}
 \label{fig3}
\end{figure}

Guided by the experimental data, we start with a model that includes a lead electrode with a hard gap on the right, a soft-gap electrode on the left, and a quantum dot in between (see supplemental materials for a full discussion of the model). Such model, however, does not generate multiple replicated spectra. To reproduce that, we need to include additional ABS in the right lead,  presumably confined within the nanowire segment underneath the superconductor. 
Good qualitative agreement is found with two ABS within the quantum dot and two ABS in the right lead, with the left lead acting as a tunneling probe (Fig.~\ref{fig2}(c)). Simulated conductance data are presented in Fig.~\ref{fig2}(d). The model exhibits multiple half-loop structures at higher bias, as well as the diagonal lines, which indeed originate from the excited states in the dot.
The horizontal resonances that bind the lowest loop are conventionally interpreted as the superconducting gap edge singularities. In our experiment this feature is observed at the scale of 0.4 meV, far below the NbTiN bulk gap. The model shows that the horizontal resonances are in fact the result of the hybridization of the lowest-energy ABS in the dot with the lowest-energy ABS in the lead. The lead ABS is not sensitive to gate $p$ therefore it appears as a horizontal resonance in the model. We also note that in practice, both devices studied in this paper likely have soft induced gaps on both sides, however essential features of the data are well captured with soft gap only on the probe side.

In order to illustrate the role of extra ABS, in Fig.~\ref{fig3} we present the results from the same basic model, in which more and more states are added to the system in subsequent panels. Fig~\ref{fig3}(a) corresponds to a single spinful ABS QD$_1$ in the quantum dot, and no ABS in the lead. It shows an Andreev loop around zero bias due to a soft gap probe (white), and a replica at the bulk gap edge (red). The Andreev loop separates the singlet regions (labeled $s_0$ and $s_2$ in panel a) and a central doublet region $d$. The three regions, which have different dot occupations ($0$ in $s_0$, 1 in $d$ and 2 in $s_2$), appear separated by discontinuities in this simulation due to the self-consistent mean-field approximation used for the interactions in the quantum dot. In Fig.~\ref{fig3}(b), a second ABS QD$_2$ is added to the quantum dot separated by $0.35$ meV from QD$_1$\cite{GharaviNanotechnology17}. At low bias, in the blue region, this yields a pair of resonances most clearly seen in the $s_0$ region.
At high bias $V>\Delta/e=2.7$ mV, in the dark-red region, additional parallel lines appear as replicas of the low bias QD$_1$ and QD$_2$ resonances.

In Fig.~\ref{fig3}(c) we have a single ABS in the dot QD$_1$ and an ABS in the lead (labeled LEAD$_1$). The latter introduces resonances that run largely parallel to the horizontal axis as in Fig.~\ref{fig2}(d). However, at the points where the lead ABS is resonant with the dot ABS the features due to QD$_1$ and LEAD$_1$ exhibit anticrossings. The lowest bias resonance transforms into a loop confined to $\pm 0.5$ meV, well below the superconducting gap. The doublet region $d$ contains more resonances than singlet regions $s_0$ and $s_2$ because ABS of different spins are not degenerate in this region.

In Fig.~\ref{fig3}(d), we again have two ABS in the lead and two in the dot, as in Fig. 2(d). Comparing with Fig.3(c), we can see additional loops forming in the low bias region, due to anticrossing of LEAD$_1$ and LEAD$_2$ with QD$_1$ and QD$_2$. The higher bias loops, as probed by the soft gap in the left electrode, show a stronger bias asymmetry in terms of peak height than the primary loop around zero bias. As already discussed, all of the low-bias features develop strong replicas due to the gap edge in the probe (red) accompanied by NDC dips (black).

\begin{figure}[ht]
  \centering
  \includegraphics[width=7cm]{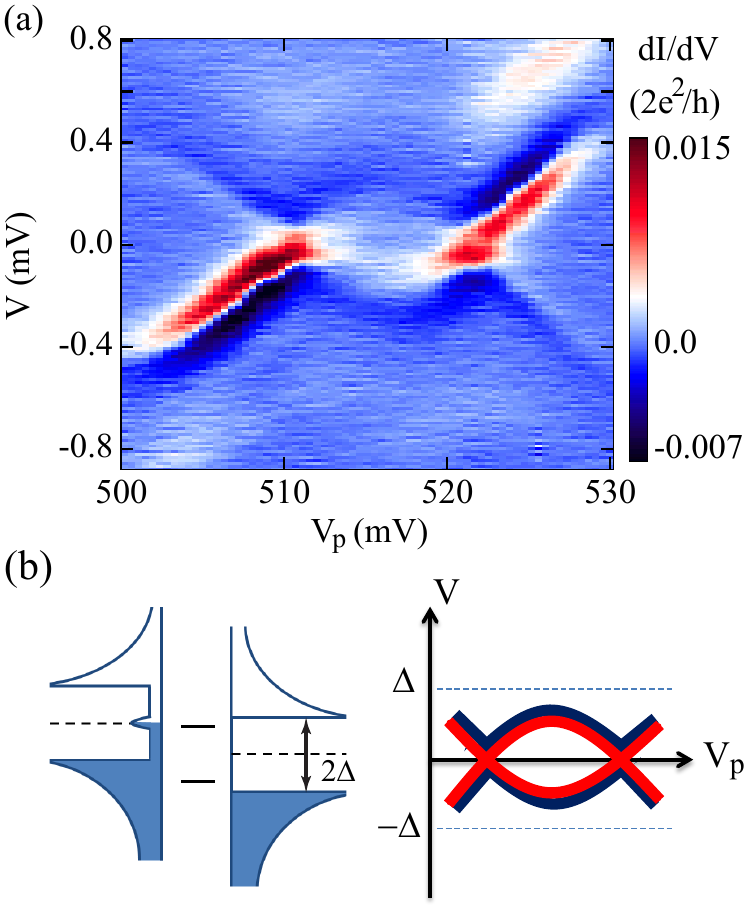}
  \caption{(a) Data in the regime similar to Fig.~\ref{fig2}(b). (b) Illustrative energy diagram with a peak in the density of states in the left probe that aligns with ABS, and produces NDC in the loop-like structure within the superconducting gap.
}
 \label{fig4}
\end{figure} 

In Fig. 4 we focus on the NDC features observed in NbTiN/InSb devices since they represent an open challenge. The unusual aspect is that NDC is observed at low bias, well within the superconducting gap (Fig.~\ref{fig4}(a)). The NDC regions trace out the loop-like Andreev resonance, at certain instances dominating over the positive differential conductance part. In differential conductance measurements, NDC often appears when two peaks in the density of states are aligned in the probe and the lead contacts. Due to increased density of states, tunneling current exhibits a peak which translates into a peak-dip structure in differential conductance. This is why NDC is often observed when tunneling from one superconducting gap edge into another (e.g. as in Fig.~\ref{fig1}(c) at $V = 2 \Delta /e$). However, NDC below the gap would require a peak in the DOS of the probe at zero bias (Fig.~\ref{fig4}(b)). Such peak is included in the model calculation in Fig.~\ref{fig2}(d). It is responsible for NDC at low bias in the model. Shifting the DOS peak in the probe to finite bias results in additional doubling of all resonant features and poorly matches the experimental data (simulation not shown).

The origin of this deduced zero-bias DOS peak, observed in several devices, is unknown at present, but it has significant implications for the interpretation of Majorana experiments done in similar devices, since MBS also manifests as a zero-bias peak. 
One can rule out Majorana as an explanation for this peak, because the subgap NDC is observed regardless of the presence of magnetic field which is a necessary ingredient for MBS. A plausible scenario is the presence of an accidental discrete zero-energy state in the probe region of the device. The local gates in that part are tuned to highly positive voltages to avoid creating additional quantum dots, and the superconducting contacts to the nanowire are highly transparent. Nevertheless, some bound states may also appear in the probe segment due to its finite size.

An important conclusion for Majorana experiments is that the tunneling probe can be more complex than a Fermi level or a textbook superconducting DOS, as confined quantum states can form in the adjacent nanowire sections, resulting in additional transport resonances. The presence of such additional resonances may complicate the interpretation of experiments aimed at detecting MBS in nanowires, and should be carefully considered. 

\textit{Acknowledgements.} We thank D. Pekker, J. Chen, P. Yu for discussions. S.M.F. is supported by NSF DMR-1743972, ONR and ARO; C.J.P. and S.M.F. by  NSF PIRE-1743717. R.A., P.S-J. and E. P. acknowledge support from the Spanish Ministry of Economy and Competitiveness through Grant Nos. FIS2015-64654-P (MINECO/FEDER), FIS2015-65706-P (MINECO/FEDER) and FIS2016-80434-P (AEI/FEDER, EU), and the Ram\'on y Cajal programme Grants RYC-2011-09345. E.J.H.L. acknowledges ERC Grant No. 716559, the Maria de Maeztu programme for Units of Excellence in R\&D (MDM-2014--0377) and the Ram\'on y Cajal programme (RYC-2015-17973).

\bibliographystyle{apsrev4-1}
\bibliography{Ref.bib}

%merlin.mbs apsrev4-1.bst 2010-07-25 4.21a (PWD, AO, DPC) hacked
%Control: key (0)
%Control: author (72) initials jnrlst
%Control: editor formatted (1) identically to author
%Control: production of article title (-1) disabled
%Control: page (0) single
%Control: year (1) truncated
%Control: production of eprint (0) enabled
\begin{thebibliography}{30}%
\makeatletter
\providecommand \@ifxundefined [1]{%
 \@ifx{#1\undefined}
}%
\providecommand \@ifnum [1]{%
 \ifnum #1\expandafter \@firstoftwo
 \else \expandafter \@secondoftwo
 \fi
}%
\providecommand \@ifx [1]{%
 \ifx #1\expandafter \@firstoftwo
 \else \expandafter \@secondoftwo
 \fi
}%
\providecommand \natexlab [1]{#1}%
\providecommand \enquote  [1]{``#1''}%
\providecommand \bibnamefont  [1]{#1}%
\providecommand \bibfnamefont [1]{#1}%
\providecommand \citenamefont [1]{#1}%
\providecommand \href@noop [0]{\@secondoftwo}%
\providecommand \href [0]{\begingroup \@sanitize@url \@href}%
\providecommand \@href[1]{\@@startlink{#1}\@@href}%
\providecommand \@@href[1]{\endgroup#1\@@endlink}%
\providecommand \@sanitize@url [0]{\catcode `\\12\catcode `\$12\catcode
  `\&12\catcode `\#12\catcode `\^12\catcode `\_12\catcode `\%12\relax}%
\providecommand \@@startlink[1]{}%
\providecommand \@@endlink[0]{}%
\providecommand \url  [0]{\begingroup\@sanitize@url \@url }%
\providecommand \@url [1]{\endgroup\@href {#1}{\urlprefix }}%
\providecommand \urlprefix  [0]{URL }%
\providecommand \Eprint [0]{\href }%
\providecommand \doibase [0]{http://dx.doi.org/}%
\providecommand \selectlanguage [0]{\@gobble}%
\providecommand \bibinfo  [0]{\@secondoftwo}%
\providecommand \bibfield  [0]{\@secondoftwo}%
\providecommand \translation [1]{[#1]}%
\providecommand \BibitemOpen [0]{}%
\providecommand \bibitemStop [0]{}%
\providecommand \bibitemNoStop [0]{.\EOS\space}%
\providecommand \EOS [0]{\spacefactor3000\relax}%
\providecommand \BibitemShut  [1]{\csname bibitem#1\endcsname}%
\let\auto@bib@innerbib\@empty
%</preamble>
\bibitem [{\citenamefont {Beenakker}(2013)}]{Beenakker:11}%
  \BibitemOpen
  \bibfield  {author} {\bibinfo {author} {\bibfnamefont {C.}~\bibnamefont
  {Beenakker}},\ }\href@noop {} {\bibfield  {journal} {\bibinfo  {journal}
  {Annu. Rev. Cond. Mat. Phys.}\ }\textbf {\bibinfo {volume} {4}},\ \bibinfo
  {pages} {113} (\bibinfo {year} {2013})}\BibitemShut {NoStop}%
\bibitem [{\citenamefont {Alicea}(2012)}]{Alicea:RPP12}%
  \BibitemOpen
  \bibfield  {author} {\bibinfo {author} {\bibfnamefont {J.}~\bibnamefont
  {Alicea}},\ }\href {http://stacks.iop.org/0034-4885/75/i=7/a=076501}
  {\bibfield  {journal} {\bibinfo  {journal} {Rep. Prog. Phys.}\ }\textbf
  {\bibinfo {volume} {75}},\ \bibinfo {pages} {076501} (\bibinfo {year}
  {2012})}\BibitemShut {NoStop}%
\bibitem [{\citenamefont {Elliott}\ and\ \citenamefont
  {Franz}(2015)}]{RevModPhys.87.137}%
  \BibitemOpen
  \bibfield  {author} {\bibinfo {author} {\bibfnamefont {S.~R.}\ \bibnamefont
  {Elliott}}\ and\ \bibinfo {author} {\bibfnamefont {M.}~\bibnamefont
  {Franz}},\ }\href {\doibase 10.1103/RevModPhys.87.137} {\bibfield  {journal}
  {\bibinfo  {journal} {Rev. Mod. Phys.}\ }\textbf {\bibinfo {volume} {87}},\
  \bibinfo {pages} {137} (\bibinfo {year} {2015})}\BibitemShut {NoStop}%
\bibitem [{\citenamefont {Aguado}(2017)}]{Aguadoreview17}%
  \BibitemOpen
  \bibfield  {author} {\bibinfo {author} {\bibfnamefont {R.}~\bibnamefont
  {Aguado}},\ }\href@noop {} {\bibfield  {journal} {\bibinfo  {journal} {La
  Rivista del Nuovo Cimento}\ }\textbf {\bibinfo {volume} {40}},\ \bibinfo
  {pages} {523} (\bibinfo {year} {2017})}\BibitemShut {NoStop}%
\bibitem [{\citenamefont {Sand-Jespersen}\ \emph {et~al.}(2007)\citenamefont
  {Sand-Jespersen}, \citenamefont {Paaske}, \citenamefont {Andersen},
  \citenamefont {Grove-Rasmussen} \emph {et~al.}}]{GroveRasmussenPRL07}%
  \BibitemOpen
  \bibfield  {author} {\bibinfo {author} {\bibfnamefont {T.}~\bibnamefont
  {Sand-Jespersen}}, \bibinfo {author} {\bibfnamefont {J.}~\bibnamefont
  {Paaske}}, \bibinfo {author} {\bibfnamefont {B.~M.}\ \bibnamefont
  {Andersen}}, \bibinfo {author} {\bibfnamefont {K.}~\bibnamefont
  {Grove-Rasmussen}},  \emph {et~al.},\ }\href {\doibase
  10.1103/PhysRevLett.99.126603} {\bibfield  {journal} {\bibinfo  {journal}
  {Phys. Rev. Lett.}\ }\textbf {\bibinfo {volume} {99}},\ \bibinfo {pages}
  {126603} (\bibinfo {year} {2007})}\BibitemShut {NoStop}%
\bibitem [{\citenamefont {Eichler}\ \emph {et~al.}(2007)\citenamefont
  {Eichler}, \citenamefont {Weiss}, \citenamefont {Oberholzer}, \citenamefont
  {Sch\"onenberger}, \citenamefont {Levy~Yeyati}, \citenamefont {Cuevas},\ and\
  \citenamefont {Mart\'{\i}n-Rodero}}]{EichlerPRL07}%
  \BibitemOpen
  \bibfield  {author} {\bibinfo {author} {\bibfnamefont {A.}~\bibnamefont
  {Eichler}}, \bibinfo {author} {\bibfnamefont {M.}~\bibnamefont {Weiss}},
  \bibinfo {author} {\bibfnamefont {S.}~\bibnamefont {Oberholzer}}, \bibinfo
  {author} {\bibfnamefont {C.}~\bibnamefont {Sch\"onenberger}}, \bibinfo
  {author} {\bibfnamefont {A.}~\bibnamefont {Levy~Yeyati}}, \bibinfo {author}
  {\bibfnamefont {J.~C.}\ \bibnamefont {Cuevas}}, \ and\ \bibinfo {author}
  {\bibfnamefont {A.}~\bibnamefont {Mart\'{\i}n-Rodero}},\ }\href {\doibase
  10.1103/PhysRevLett.99.126602} {\bibfield  {journal} {\bibinfo  {journal}
  {Phys. Rev. Lett.}\ }\textbf {\bibinfo {volume} {99}},\ \bibinfo {pages}
  {126602} (\bibinfo {year} {2007})}\BibitemShut {NoStop}%
\bibitem [{\citenamefont {Deacon}\ \emph {et~al.}(2010)\citenamefont {Deacon},
  \citenamefont {Tanaka}, \citenamefont {Oiwa}, \citenamefont {Sakano},
  \citenamefont {Yoshida}, \citenamefont {Shibata}, \citenamefont {Hirakawa},\
  and\ \citenamefont {Tarucha}}]{ABS_Tarucha_PRL}%
  \BibitemOpen
  \bibfield  {author} {\bibinfo {author} {\bibfnamefont {R.~S.}\ \bibnamefont
  {Deacon}}, \bibinfo {author} {\bibfnamefont {Y.}~\bibnamefont {Tanaka}},
  \bibinfo {author} {\bibfnamefont {A.}~\bibnamefont {Oiwa}}, \bibinfo {author}
  {\bibfnamefont {R.}~\bibnamefont {Sakano}}, \bibinfo {author} {\bibfnamefont
  {K.}~\bibnamefont {Yoshida}}, \bibinfo {author} {\bibfnamefont
  {K.}~\bibnamefont {Shibata}}, \bibinfo {author} {\bibfnamefont
  {K.}~\bibnamefont {Hirakawa}}, \ and\ \bibinfo {author} {\bibfnamefont
  {S.}~\bibnamefont {Tarucha}},\ }\href@noop {} {\bibfield  {journal} {\bibinfo
   {journal} {Phys. Rev. Lett.}\ }\textbf {\bibinfo {volume} {104}},\ \bibinfo
  {pages} {076805} (\bibinfo {year} {2010})}\BibitemShut {NoStop}%
\bibitem [{\citenamefont {Pillet}\ \emph {et~al.}(2010)\citenamefont {Pillet},
  \citenamefont {Quay}, \citenamefont {Morfin}, \citenamefont {Bena},
  \citenamefont {Levy~Yeyati}, ,\ and\ \citenamefont {Joyez}}]{first_ABS}%
  \BibitemOpen
  \bibfield  {author} {\bibinfo {author} {\bibfnamefont {J.-D.}\ \bibnamefont
  {Pillet}}, \bibinfo {author} {\bibfnamefont {C.~H.~L.}\ \bibnamefont {Quay}},
  \bibinfo {author} {\bibfnamefont {P.}~\bibnamefont {Morfin}}, \bibinfo
  {author} {\bibfnamefont {C.}~\bibnamefont {Bena}}, \bibinfo {author}
  {\bibfnamefont {A.}~\bibnamefont {Levy~Yeyati}}, , \ and\ \bibinfo {author}
  {\bibfnamefont {P.}~\bibnamefont {Joyez}},\ }\href@noop {} {\bibfield
  {journal} {\bibinfo  {journal} {Nature Phys.}\ }\textbf {\bibinfo {volume}
  {6}},\ \bibinfo {pages} {965} (\bibinfo {year} {2010})}\BibitemShut {NoStop}%
\bibitem [{\citenamefont {Dirks}\ \emph {et~al.}(2011)\citenamefont {Dirks},
  \citenamefont {Hughes}, \citenamefont {Lal}, \citenamefont {Uchoa},
  \citenamefont {Chen}, \citenamefont {Chialvo}, \citenamefont {Goldbart},\
  and\ \citenamefont {Mason}}]{ABS_Nadya}%
  \BibitemOpen
  \bibfield  {author} {\bibinfo {author} {\bibfnamefont {T.}~\bibnamefont
  {Dirks}}, \bibinfo {author} {\bibfnamefont {T.~L.}\ \bibnamefont {Hughes}},
  \bibinfo {author} {\bibfnamefont {S.}~\bibnamefont {Lal}}, \bibinfo {author}
  {\bibfnamefont {B.}~\bibnamefont {Uchoa}}, \bibinfo {author} {\bibfnamefont
  {Y.-F.}\ \bibnamefont {Chen}}, \bibinfo {author} {\bibfnamefont
  {C.}~\bibnamefont {Chialvo}}, \bibinfo {author} {\bibfnamefont {P.~M.}\
  \bibnamefont {Goldbart}}, \ and\ \bibinfo {author} {\bibfnamefont
  {N.}~\bibnamefont {Mason}},\ }\href@noop {} {\bibfield  {journal} {\bibinfo
  {journal} {Nature Phys.}\ }\textbf {\bibinfo {volume} {7}},\ \bibinfo {pages}
  {386} (\bibinfo {year} {2011})}\BibitemShut {NoStop}%
\bibitem [{\citenamefont {Chang}\ \emph {et~al.}(2013)\citenamefont {Chang},
  \citenamefont {Manucharyan}, \citenamefont {Jespersen}, \citenamefont
  {Nyg\aa{}rd},\ and\ \citenamefont {Marcus}}]{changPRL2013}%
  \BibitemOpen
  \bibfield  {author} {\bibinfo {author} {\bibfnamefont {W.}~\bibnamefont
  {Chang}}, \bibinfo {author} {\bibfnamefont {V.~E.}\ \bibnamefont
  {Manucharyan}}, \bibinfo {author} {\bibfnamefont {T.~S.}\ \bibnamefont
  {Jespersen}}, \bibinfo {author} {\bibfnamefont {J.}~\bibnamefont
  {Nyg\aa{}rd}}, \ and\ \bibinfo {author} {\bibfnamefont {C.~M.}\ \bibnamefont
  {Marcus}},\ }\href@noop {} {\bibfield  {journal} {\bibinfo  {journal} {Phys.
  Rev. Lett.}\ }\textbf {\bibinfo {volume} {110}},\ \bibinfo {pages} {217005}
  (\bibinfo {year} {2013})}\BibitemShut {NoStop}%
\bibitem [{\citenamefont {Bretheau}\ \emph {et~al.}(2013)\citenamefont
  {Bretheau}, \citenamefont {Girit}, \citenamefont {Pothier}, \citenamefont
  {Esteve},\ and\ \citenamefont {Urbina}}]{bretheauNature13}%
  \BibitemOpen
  \bibfield  {author} {\bibinfo {author} {\bibfnamefont {L.}~\bibnamefont
  {Bretheau}}, \bibinfo {author} {\bibfnamefont {{\c{C}}.}~\bibnamefont
  {Girit}}, \bibinfo {author} {\bibfnamefont {H.}~\bibnamefont {Pothier}},
  \bibinfo {author} {\bibfnamefont {D.}~\bibnamefont {Esteve}}, \ and\ \bibinfo
  {author} {\bibfnamefont {C.}~\bibnamefont {Urbina}},\ }\href@noop {}
  {\bibfield  {journal} {\bibinfo  {journal} {Nature}\ }\textbf {\bibinfo
  {volume} {499}},\ \bibinfo {pages} {312} (\bibinfo {year}
  {2013})}\BibitemShut {NoStop}%
\bibitem [{\citenamefont {Lee}\ \emph {et~al.}(2014)\citenamefont {Lee},
  \citenamefont {Jiang}, \citenamefont {Houzet}, \citenamefont {Aguado},
  \citenamefont {Lieber},\ and\ \citenamefont
  {De~Franceschi}}]{LeeNatnano2014}%
  \BibitemOpen
  \bibfield  {author} {\bibinfo {author} {\bibfnamefont {E.~J.~H.}\
  \bibnamefont {Lee}}, \bibinfo {author} {\bibfnamefont {X.}~\bibnamefont
  {Jiang}}, \bibinfo {author} {\bibfnamefont {M.}~\bibnamefont {Houzet}},
  \bibinfo {author} {\bibfnamefont {R.}~\bibnamefont {Aguado}}, \bibinfo
  {author} {\bibfnamefont {C.~M.}\ \bibnamefont {Lieber}}, \ and\ \bibinfo
  {author} {\bibfnamefont {S.}~\bibnamefont {De~Franceschi}},\ }\href@noop {}
  {\bibfield  {journal} {\bibinfo  {journal} {Nature nanotechnology}\ }\textbf
  {\bibinfo {volume} {9}},\ \bibinfo {pages} {79} (\bibinfo {year}
  {2014})}\BibitemShut {NoStop}%
\bibitem [{\citenamefont {Liu}\ \emph {et~al.}(2017)\citenamefont {Liu},
  \citenamefont {Sau}, \citenamefont {Stanescu},\ and\ \citenamefont
  {Das~Sarma}}]{LiuPRB17}%
  \BibitemOpen
  \bibfield  {author} {\bibinfo {author} {\bibfnamefont {C.-X.}\ \bibnamefont
  {Liu}}, \bibinfo {author} {\bibfnamefont {J.~D.}\ \bibnamefont {Sau}},
  \bibinfo {author} {\bibfnamefont {T.~D.}\ \bibnamefont {Stanescu}}, \ and\
  \bibinfo {author} {\bibfnamefont {S.}~\bibnamefont {Das~Sarma}},\ }\href
  {\doibase 10.1103/PhysRevB.96.075161} {\bibfield  {journal} {\bibinfo
  {journal} {Phys. Rev. B}\ }\textbf {\bibinfo {volume} {96}},\ \bibinfo
  {pages} {075161} (\bibinfo {year} {2017})}\BibitemShut {NoStop}%
\bibitem [{\citenamefont {Yazdani}\ \emph {et~al.}(1997)\citenamefont
  {Yazdani}, \citenamefont {Jones}, \citenamefont {Lutz}, \citenamefont
  {Crommie},\ and\ \citenamefont {Eigler}}]{Yazdani1997}%
  \BibitemOpen
  \bibfield  {author} {\bibinfo {author} {\bibfnamefont {A.}~\bibnamefont
  {Yazdani}}, \bibinfo {author} {\bibfnamefont {B.~A.}\ \bibnamefont {Jones}},
  \bibinfo {author} {\bibfnamefont {C.~P.}\ \bibnamefont {Lutz}}, \bibinfo
  {author} {\bibfnamefont {M.~F.}\ \bibnamefont {Crommie}}, \ and\ \bibinfo
  {author} {\bibfnamefont {D.~M.}\ \bibnamefont {Eigler}},\ }\href {\doibase
  10.1126/science.275.5307.1767} {\bibfield  {journal} {\bibinfo  {journal}
  {Science}\ }\textbf {\bibinfo {volume} {275}},\ \bibinfo {pages} {1767}
  (\bibinfo {year} {1997})}\BibitemShut {NoStop}%
\bibitem [{\citenamefont {Ruby}\ \emph
  {et~al.}(2015{\natexlab{a}})\citenamefont {Ruby}, \citenamefont {Pientka},
  \citenamefont {Peng}, \citenamefont {von Oppen}, \citenamefont {Heinrich},\
  and\ \citenamefont {Franke}}]{RubyPRL15}%
  \BibitemOpen
  \bibfield  {author} {\bibinfo {author} {\bibfnamefont {M.}~\bibnamefont
  {Ruby}}, \bibinfo {author} {\bibfnamefont {F.}~\bibnamefont {Pientka}},
  \bibinfo {author} {\bibfnamefont {Y.}~\bibnamefont {Peng}}, \bibinfo {author}
  {\bibfnamefont {F.}~\bibnamefont {von Oppen}}, \bibinfo {author}
  {\bibfnamefont {B.~W.}\ \bibnamefont {Heinrich}}, \ and\ \bibinfo {author}
  {\bibfnamefont {K.~J.}\ \bibnamefont {Franke}},\ }\href {\doibase
  10.1103/PhysRevLett.115.087001} {\bibfield  {journal} {\bibinfo  {journal}
  {Phys. Rev. Lett.}\ }\textbf {\bibinfo {volume} {115}},\ \bibinfo {pages}
  {087001} (\bibinfo {year} {2015}{\natexlab{a}})}\BibitemShut {NoStop}%
\bibitem [{\citenamefont {Mourik}\ \emph {et~al.}(2012)\citenamefont {Mourik},
  \citenamefont {Zuo}, \citenamefont {Frolov}, \citenamefont {Plissard},
  \citenamefont {Bakkers},\ and\ \citenamefont
  {Kouwenhoven}}]{MourikScience2012}%
  \BibitemOpen
  \bibfield  {author} {\bibinfo {author} {\bibfnamefont {V.}~\bibnamefont
  {Mourik}}, \bibinfo {author} {\bibfnamefont {K.}~\bibnamefont {Zuo}},
  \bibinfo {author} {\bibfnamefont {S.~M.}\ \bibnamefont {Frolov}}, \bibinfo
  {author} {\bibfnamefont {S.~R.}\ \bibnamefont {Plissard}}, \bibinfo {author}
  {\bibfnamefont {E.~P. A.~M.}\ \bibnamefont {Bakkers}}, \ and\ \bibinfo
  {author} {\bibfnamefont {L.~P.}\ \bibnamefont {Kouwenhoven}},\ }\href@noop {}
  {\bibfield  {journal} {\bibinfo  {journal} {Science}\ }\textbf {\bibinfo
  {volume} {336}},\ \bibinfo {pages} {1003} (\bibinfo {year}
  {2012})}\BibitemShut {NoStop}%
\bibitem [{\citenamefont {Das}\ \emph {et~al.}(2012)\citenamefont {Das},
  \citenamefont {Ronen}, \citenamefont {Most}, \citenamefont {Oreg},
  \citenamefont {Heiblum},\ and\ \citenamefont {Shtrikman}}]{DasNaturePhy2012}%
  \BibitemOpen
  \bibfield  {author} {\bibinfo {author} {\bibfnamefont {A.}~\bibnamefont
  {Das}}, \bibinfo {author} {\bibfnamefont {Y.}~\bibnamefont {Ronen}}, \bibinfo
  {author} {\bibfnamefont {Y.}~\bibnamefont {Most}}, \bibinfo {author}
  {\bibfnamefont {Y.}~\bibnamefont {Oreg}}, \bibinfo {author} {\bibfnamefont
  {M.}~\bibnamefont {Heiblum}}, \ and\ \bibinfo {author} {\bibfnamefont
  {H.}~\bibnamefont {Shtrikman}},\ }\href@noop {} {\bibfield  {journal}
  {\bibinfo  {journal} {Nature Phys.}\ }\textbf {\bibinfo {volume} {8}},\
  \bibinfo {pages} {887} (\bibinfo {year} {2012})}\BibitemShut {NoStop}%
\bibitem [{\citenamefont {Deng}\ \emph {et~al.}(2016)\citenamefont {Deng},
  \citenamefont {Vaitiekenas}, \citenamefont {Hansen}, \citenamefont {Danon},
  \citenamefont {Leijnse}, \citenamefont {Flensberg}, \citenamefont {Nyg{\r
  a}rd}, \citenamefont {Krogstrup},\ and\ \citenamefont
  {Marcus}}]{DengScience2016}%
  \BibitemOpen
  \bibfield  {author} {\bibinfo {author} {\bibfnamefont {M.~T.}\ \bibnamefont
  {Deng}}, \bibinfo {author} {\bibfnamefont {S.}~\bibnamefont {Vaitiekenas}},
  \bibinfo {author} {\bibfnamefont {E.~B.}\ \bibnamefont {Hansen}}, \bibinfo
  {author} {\bibfnamefont {J.}~\bibnamefont {Danon}}, \bibinfo {author}
  {\bibfnamefont {M.}~\bibnamefont {Leijnse}}, \bibinfo {author} {\bibfnamefont
  {K.}~\bibnamefont {Flensberg}}, \bibinfo {author} {\bibfnamefont
  {J.}~\bibnamefont {Nyg{\r a}rd}}, \bibinfo {author} {\bibfnamefont
  {P.}~\bibnamefont {Krogstrup}}, \ and\ \bibinfo {author} {\bibfnamefont
  {C.~M.}\ \bibnamefont {Marcus}},\ }\href {\doibase 10.1126/science.aaf3961}
  {\bibfield  {journal} {\bibinfo  {journal} {Science}\ }\textbf {\bibinfo
  {volume} {354}},\ \bibinfo {pages} {1557} (\bibinfo {year}
  {2016})}\BibitemShut {NoStop}%
\bibitem [{\citenamefont {Chen}\ \emph {et~al.}(2017)\citenamefont {Chen},
  \citenamefont {Yu}, \citenamefont {Stenger}, \citenamefont {Hocevar},
  \citenamefont {Car}, \citenamefont {Plissard}, \citenamefont {Bakkers},
  \citenamefont {Stanescu},\ and\ \citenamefont {Frolov}}]{Chen2017}%
  \BibitemOpen
  \bibfield  {author} {\bibinfo {author} {\bibfnamefont {J.}~\bibnamefont
  {Chen}}, \bibinfo {author} {\bibfnamefont {P.}~\bibnamefont {Yu}}, \bibinfo
  {author} {\bibfnamefont {J.}~\bibnamefont {Stenger}}, \bibinfo {author}
  {\bibfnamefont {M.}~\bibnamefont {Hocevar}}, \bibinfo {author} {\bibfnamefont
  {D.}~\bibnamefont {Car}}, \bibinfo {author} {\bibfnamefont {S.~R.}\
  \bibnamefont {Plissard}}, \bibinfo {author} {\bibfnamefont {E.~P. A.~M.}\
  \bibnamefont {Bakkers}}, \bibinfo {author} {\bibfnamefont {T.~D.}\
  \bibnamefont {Stanescu}}, \ and\ \bibinfo {author} {\bibfnamefont {S.~M.}\
  \bibnamefont {Frolov}},\ }\href@noop {} {\bibfield  {journal} {\bibinfo
  {journal} {Science Advances}\ }\textbf {\bibinfo {volume} {3}} (\bibinfo
  {year} {2017})}\BibitemShut {NoStop}%
\bibitem [{\citenamefont {Zhang}\ \emph {et~al.}(2018)\citenamefont {Zhang},
  \citenamefont {Liu}, \citenamefont {Gazibegovic}, \citenamefont {Xu},
  \citenamefont {Logan}, \citenamefont {Wang}, \citenamefont {van Loo},
  \citenamefont {Bommer} \emph {et~al.}}]{quantized_majorana}%
  \BibitemOpen
  \bibfield  {author} {\bibinfo {author} {\bibfnamefont {H.}~\bibnamefont
  {Zhang}}, \bibinfo {author} {\bibfnamefont {C.-X.}\ \bibnamefont {Liu}},
  \bibinfo {author} {\bibfnamefont {S.}~\bibnamefont {Gazibegovic}}, \bibinfo
  {author} {\bibfnamefont {D.}~\bibnamefont {Xu}}, \bibinfo {author}
  {\bibfnamefont {J.~A.}\ \bibnamefont {Logan}}, \bibinfo {author}
  {\bibfnamefont {G.}~\bibnamefont {Wang}}, \bibinfo {author} {\bibfnamefont
  {N.}~\bibnamefont {van Loo}}, \bibinfo {author} {\bibfnamefont {J.~D.~S.}\
  \bibnamefont {Bommer}},  \emph {et~al.},\ }\href@noop {} {\bibfield
  {journal} {\bibinfo  {journal} {Nature}\ }\textbf {\bibinfo {volume} {556}},\
  \bibinfo {pages} {74} (\bibinfo {year} {2018})}\BibitemShut {NoStop}%
\bibitem [{\citenamefont {G{\"u}l}\ \emph {et~al.}(2018)\citenamefont
  {G{\"u}l}, \citenamefont {Zhang}, \citenamefont {Bommer}, \citenamefont
  {de~Moor}, \citenamefont {Car}, \citenamefont {Plissard}, \citenamefont
  {Bakkers}, \citenamefont {Geresdi}, \citenamefont {Watanabe}, \citenamefont
  {Taniguchi} \emph {et~al.}}]{GulNatNano18}%
  \BibitemOpen
  \bibfield  {author} {\bibinfo {author} {\bibfnamefont {{\"O}.}~\bibnamefont
  {G{\"u}l}}, \bibinfo {author} {\bibfnamefont {H.}~\bibnamefont {Zhang}},
  \bibinfo {author} {\bibfnamefont {J.~D.}\ \bibnamefont {Bommer}}, \bibinfo
  {author} {\bibfnamefont {M.~W.}\ \bibnamefont {de~Moor}}, \bibinfo {author}
  {\bibfnamefont {D.}~\bibnamefont {Car}}, \bibinfo {author} {\bibfnamefont
  {S.~R.}\ \bibnamefont {Plissard}}, \bibinfo {author} {\bibfnamefont {E.~P.}\
  \bibnamefont {Bakkers}}, \bibinfo {author} {\bibfnamefont {A.}~\bibnamefont
  {Geresdi}}, \bibinfo {author} {\bibfnamefont {K.}~\bibnamefont {Watanabe}},
  \bibinfo {author} {\bibfnamefont {T.}~\bibnamefont {Taniguchi}},  \emph
  {et~al.},\ }\href@noop {} {\bibfield  {journal} {\bibinfo  {journal} {Nature
  nanotechnology}\ }\textbf {\bibinfo {volume} {13}},\ \bibinfo {pages} {192}
  (\bibinfo {year} {2018})}\BibitemShut {NoStop}%
\bibitem [{\citenamefont {Plissard}\ \emph {et~al.}(2012)\citenamefont
  {Plissard}, \citenamefont {Slapak}, \citenamefont {Verheijen}, \citenamefont
  {Hocevar}, \citenamefont {Immink}, \citenamefont {van Weperen}, \citenamefont
  {Nadj-Perge}, \citenamefont {Frolov}, \citenamefont {Kouwenhoven},\ and\
  \citenamefont {Bakkers}}]{PlissardNanoLett12}%
  \BibitemOpen
  \bibfield  {author} {\bibinfo {author} {\bibfnamefont {S.~R.}\ \bibnamefont
  {Plissard}}, \bibinfo {author} {\bibfnamefont {D.~R.}\ \bibnamefont
  {Slapak}}, \bibinfo {author} {\bibfnamefont {M.~A.}\ \bibnamefont
  {Verheijen}}, \bibinfo {author} {\bibfnamefont {M.}~\bibnamefont {Hocevar}},
  \bibinfo {author} {\bibfnamefont {G.~W.~G.}\ \bibnamefont {Immink}}, \bibinfo
  {author} {\bibfnamefont {I.}~\bibnamefont {van Weperen}}, \bibinfo {author}
  {\bibfnamefont {S.}~\bibnamefont {Nadj-Perge}}, \bibinfo {author}
  {\bibfnamefont {S.~M.}\ \bibnamefont {Frolov}}, \bibinfo {author}
  {\bibfnamefont {L.~P.}\ \bibnamefont {Kouwenhoven}}, \ and\ \bibinfo {author}
  {\bibfnamefont {E.~P. A.~M.}\ \bibnamefont {Bakkers}},\ }\href@noop {}
  {\bibfield  {journal} {\bibinfo  {journal} {Nano Letters}\ }\textbf {\bibinfo
  {volume} {12}},\ \bibinfo {pages} {1794} (\bibinfo {year}
  {2012})}\BibitemShut {NoStop}%
\bibitem [{\citenamefont {Gazibegovic}\ \emph {et~al.}(2017)\citenamefont
  {Gazibegovic}, \citenamefont {Car}, \citenamefont {Zhang}, \citenamefont
  {Balk}, \citenamefont {Logan}, \citenamefont {de~Moor}, \citenamefont
  {Cassidy}, \citenamefont {Schmits} \emph {et~al.}}]{gazibegovic2017epitaxy}%
  \BibitemOpen
  \bibfield  {author} {\bibinfo {author} {\bibfnamefont {S.}~\bibnamefont
  {Gazibegovic}}, \bibinfo {author} {\bibfnamefont {D.}~\bibnamefont {Car}},
  \bibinfo {author} {\bibfnamefont {H.}~\bibnamefont {Zhang}}, \bibinfo
  {author} {\bibfnamefont {S.~C.}\ \bibnamefont {Balk}}, \bibinfo {author}
  {\bibfnamefont {J.~A.}\ \bibnamefont {Logan}}, \bibinfo {author}
  {\bibfnamefont {M.~W.}\ \bibnamefont {de~Moor}}, \bibinfo {author}
  {\bibfnamefont {M.~C.}\ \bibnamefont {Cassidy}}, \bibinfo {author}
  {\bibfnamefont {R.}~\bibnamefont {Schmits}},  \emph {et~al.},\ }\href@noop {}
  {\bibfield  {journal} {\bibinfo  {journal} {Nature}\ }\textbf {\bibinfo
  {volume} {548}},\ \bibinfo {pages} {434 } (\bibinfo {year}
  {2017})}\BibitemShut {NoStop}%
\bibitem [{\citenamefont {Ruby}\ \emph
  {et~al.}(2015{\natexlab{b}})\citenamefont {Ruby}, \citenamefont {Pientka},
  \citenamefont {Peng}, \citenamefont {von Oppen}, \citenamefont {Heinrich},\
  and\ \citenamefont {Franke}}]{Pientka2015}%
  \BibitemOpen
  \bibfield  {author} {\bibinfo {author} {\bibfnamefont {M.}~\bibnamefont
  {Ruby}}, \bibinfo {author} {\bibfnamefont {F.}~\bibnamefont {Pientka}},
  \bibinfo {author} {\bibfnamefont {Y.}~\bibnamefont {Peng}}, \bibinfo {author}
  {\bibfnamefont {F.}~\bibnamefont {von Oppen}}, \bibinfo {author}
  {\bibfnamefont {B.~W.}\ \bibnamefont {Heinrich}}, \ and\ \bibinfo {author}
  {\bibfnamefont {K.~J.}\ \bibnamefont {Franke}},\ }\href {\doibase
  10.1103/PhysRevLett.115.197204} {\bibfield  {journal} {\bibinfo  {journal}
  {Phys. Rev. Lett.}\ }\textbf {\bibinfo {volume} {115}},\ \bibinfo {pages}
  {197204} (\bibinfo {year} {2015}{\natexlab{b}})}\BibitemShut {NoStop}%
\bibitem [{\citenamefont {Grove-Rasmussen}\ \emph {et~al.}(2009)\citenamefont
  {Grove-Rasmussen}, \citenamefont {J\o{}rgensen}, \citenamefont {Andersen},
  \citenamefont {Paaske}, \citenamefont {Jespersen}, \citenamefont
  {Nyg\aa{}rd}, \citenamefont {Flensberg},\ and\ \citenamefont
  {Lindelof}}]{GroveRasmussenPRB09}%
  \BibitemOpen
  \bibfield  {author} {\bibinfo {author} {\bibfnamefont {K.}~\bibnamefont
  {Grove-Rasmussen}}, \bibinfo {author} {\bibfnamefont {H.~I.}\ \bibnamefont
  {J\o{}rgensen}}, \bibinfo {author} {\bibfnamefont {B.~M.}\ \bibnamefont
  {Andersen}}, \bibinfo {author} {\bibfnamefont {J.}~\bibnamefont {Paaske}},
  \bibinfo {author} {\bibfnamefont {T.~S.}\ \bibnamefont {Jespersen}}, \bibinfo
  {author} {\bibfnamefont {J.}~\bibnamefont {Nyg\aa{}rd}}, \bibinfo {author}
  {\bibfnamefont {K.}~\bibnamefont {Flensberg}}, \ and\ \bibinfo {author}
  {\bibfnamefont {P.~E.}\ \bibnamefont {Lindelof}},\ }\href {\doibase
  10.1103/PhysRevB.79.134518} {\bibfield  {journal} {\bibinfo  {journal} {Phys.
  Rev. B}\ }\textbf {\bibinfo {volume} {79}},\ \bibinfo {pages} {134518}
  (\bibinfo {year} {2009})}\BibitemShut {NoStop}%
\bibitem [{\citenamefont {Lee}\ \emph {et~al.}(2012)\citenamefont {Lee},
  \citenamefont {Jiang}, \citenamefont {Aguado}, \citenamefont {Katsaros},
  \citenamefont {Lieber},\ and\ \citenamefont {De~Franceschi}}]{LeePRL2012}%
  \BibitemOpen
  \bibfield  {author} {\bibinfo {author} {\bibfnamefont {E.~J.~H.}\
  \bibnamefont {Lee}}, \bibinfo {author} {\bibfnamefont {X.}~\bibnamefont
  {Jiang}}, \bibinfo {author} {\bibfnamefont {R.}~\bibnamefont {Aguado}},
  \bibinfo {author} {\bibfnamefont {G.}~\bibnamefont {Katsaros}}, \bibinfo
  {author} {\bibfnamefont {C.~M.}\ \bibnamefont {Lieber}}, \ and\ \bibinfo
  {author} {\bibfnamefont {S.}~\bibnamefont {De~Franceschi}},\ }\href {\doibase
  10.1103/PhysRevLett.109.186802} {\bibfield  {journal} {\bibinfo  {journal}
  {Phys. Rev. Lett.}\ }\textbf {\bibinfo {volume} {109}},\ \bibinfo {pages}
  {186802} (\bibinfo {year} {2012})}\BibitemShut {NoStop}%
\bibitem [{\citenamefont {Su}\ \emph {et~al.}(2017)\citenamefont {Su},
  \citenamefont {Tacla}, \citenamefont {Hocevar}, \citenamefont {Car},
  \citenamefont {Plissard}, \citenamefont {Bakkers}, \citenamefont {Daley},
  \citenamefont {Pekker},\ and\ \citenamefont {Frolov}}]{su_andreevmolecule}%
  \BibitemOpen
  \bibfield  {author} {\bibinfo {author} {\bibfnamefont {Z.}~\bibnamefont
  {Su}}, \bibinfo {author} {\bibfnamefont {A.~B.}\ \bibnamefont {Tacla}},
  \bibinfo {author} {\bibfnamefont {M.}~\bibnamefont {Hocevar}}, \bibinfo
  {author} {\bibfnamefont {D.}~\bibnamefont {Car}}, \bibinfo {author}
  {\bibfnamefont {S.~R.}\ \bibnamefont {Plissard}}, \bibinfo {author}
  {\bibfnamefont {E.~P. A.~M.}\ \bibnamefont {Bakkers}}, \bibinfo {author}
  {\bibfnamefont {A.~J.}\ \bibnamefont {Daley}}, \bibinfo {author}
  {\bibfnamefont {D.}~\bibnamefont {Pekker}}, \ and\ \bibinfo {author}
  {\bibfnamefont {S.~M.}\ \bibnamefont {Frolov}},\ }\href@noop {} {\bibfield
  {journal} {\bibinfo  {journal} {Nature Communications}\ }\textbf {\bibinfo
  {volume} {8}},\ \bibinfo {pages} {585} (\bibinfo {year} {2017})}\BibitemShut
  {NoStop}%
\bibitem [{\citenamefont {Gharavi}\ \emph {et~al.}(2017)\citenamefont
  {Gharavi}, \citenamefont {Holloway}, \citenamefont {LaPierre},\ and\
  \citenamefont {Baugh}}]{GharaviNanotechnology17}%
  \BibitemOpen
  \bibfield  {author} {\bibinfo {author} {\bibfnamefont {K.}~\bibnamefont
  {Gharavi}}, \bibinfo {author} {\bibfnamefont {G.~W.}\ \bibnamefont
  {Holloway}}, \bibinfo {author} {\bibfnamefont {R.~R.}\ \bibnamefont
  {LaPierre}}, \ and\ \bibinfo {author} {\bibfnamefont {J.}~\bibnamefont
  {Baugh}},\ }\href@noop {} {\bibfield  {journal} {\bibinfo  {journal}
  {Nanotechnology}\ }\textbf {\bibinfo {volume} {28}},\ \bibinfo {pages}
  {085202} (\bibinfo {year} {2017})}\BibitemShut {NoStop}%
\bibitem [{\citenamefont {Martin-Rodero}\ and\ \citenamefont
  {Levy~Yeyati}(2012)}]{Martin-Rodero12}%
  \BibitemOpen
  \bibfield  {author} {\bibinfo {author} {\bibfnamefont {A.}~\bibnamefont
  {Martin-Rodero}}\ and\ \bibinfo {author} {\bibfnamefont {A.}~\bibnamefont
  {Levy~Yeyati}},\ }\href@noop {} {\bibfield  {journal} {\bibinfo  {journal}
  {Journal of Physics: Condensed Matter}\ }\textbf {\bibinfo {volume} {24}},\
  \bibinfo {pages} {385303} (\bibinfo {year} {2012})}\BibitemShut {NoStop}%
\bibitem [{\citenamefont {Grove-Rasmussen}\ \emph {et~al.}(2017)\citenamefont
  {Grove-Rasmussen}, \citenamefont {Steffensen}, \citenamefont {Jellinggaard},
  \citenamefont {Madsen}, \citenamefont {Zitko}, \citenamefont {Paaske},\ and\
  \citenamefont {Nygard}}]{Grove-Rasmussen17}%
  \BibitemOpen
  \bibfield  {author} {\bibinfo {author} {\bibfnamefont {K.}~\bibnamefont
  {Grove-Rasmussen}}, \bibinfo {author} {\bibfnamefont {G.}~\bibnamefont
  {Steffensen}}, \bibinfo {author} {\bibfnamefont {A.}~\bibnamefont
  {Jellinggaard}}, \bibinfo {author} {\bibfnamefont {M.~H.}\ \bibnamefont
  {Madsen}}, \bibinfo {author} {\bibfnamefont {R.}~\bibnamefont {Zitko}},
  \bibinfo {author} {\bibfnamefont {J.}~\bibnamefont {Paaske}}, \ and\ \bibinfo
  {author} {\bibfnamefont {J.}~\bibnamefont {Nygard}},\ }\href@noop {}
  {\bibfield  {journal} {\bibinfo  {journal} {arXiv preprint arXiv:1711.06081}\
  } (\bibinfo {year} {2017})}\BibitemShut {NoStop}%
\end{thebibliography}%

\newpage
\section{SUPPLEMENTARY MATERIALS}
\setcounter{figure}{0}
\renewcommand{\thefigure}{S\arabic{figure}}

\textbf{Theoretical model.} This section introduces the model of the form $H=H_d+H_R+V$ for the quantum dot coupled to the right hard-gap superconducting electrode. The left electrode is treated as a superconducting tunnel probe and is discussed later. The quantum dot is modeled as a multiorbital Anderson model, with levels $\epsilon^D_i$, of the form 
\begin{equation}
H_d=\sum_{i\sigma}\epsilon^D_i d_{i\sigma}^\dagger d_{i\sigma} + U\sum_in_{i\uparrow}n_{i\downarrow}
%U(\langle n_\uparrow\rangle n_\downarrow + \langle n_\downarrow\rangle n_\uparrow),
\end{equation}
where $n_{i\sigma}=d^\dagger_{i\sigma}d_{i\sigma}$. 
As usual, we model the coupling to the right superconductor by a self energy $\Sigma(\omega)$ that captures the induced  pairing $\Gamma_S$ in the dot resulting from the parent gap $\Delta_p$. In the Bogoliubov-de Gennes basis $D_i=\left(d_{i\uparrow},d_{i\downarrow},d^\dagger_{i\uparrow},d^\dagger_{i\downarrow}\right)$, this dot self-energy reads
\begin{equation}
\label{Sigma}
\Sigma_R(\omega) = \frac{1}{2}\frac{\Gamma_S}{\sqrt{\Delta_p^2-\omega^2}}\sum_{i\sigma} D^\dagger_i\left(\begin{array}{cccc}
-\omega & 0 & 0 & -\Delta_p \\
0 & -\omega & \Delta_p & 0 \\
0 & \Delta_p & -\omega & 0 \\
-\Delta_p & 0 & 0 & -\omega
\end{array}\right)D_i
\end{equation}
For the physics discussed here, it is a good approximation to solve the above superconducting Anderson Hamiltonian in a Hartree-Fock mean field approximation, hence neglecting Kondo correlations, for a discussion see e.g. Ref. [\onlinecite{Martin-Rodero12}].
Clearly, this standard model of a proximitized quantum dot does not capture in full the experimental phenomenology (in particular the replicas described in the main text), so we consider the possibility that the superconducting electrode contains additional ABSs (owing to its finite length). The ABSs are described by the term
\begin{equation}
H_R=\frac{1}{2}\sum_{j\sigma}\epsilon^L_i\,c_{i\sigma} c_{i\bar\sigma} + \mathrm{h.c.},
\end{equation} 
which is essentially a generalization of the so-called zero-bandwidth model to include more than one quasiparticle excitation in the superconducting lead. Note that for the case of a two orbital Anderson model coupled to a single quasiparticle excitation in the superconducting lead (see Fig. 3c) our model recovers the double dot model of Ref. [\onlinecite{Grove-Rasmussen17}].

\begin{figure} 
  \centering
  \includegraphics[width=7.5cm]{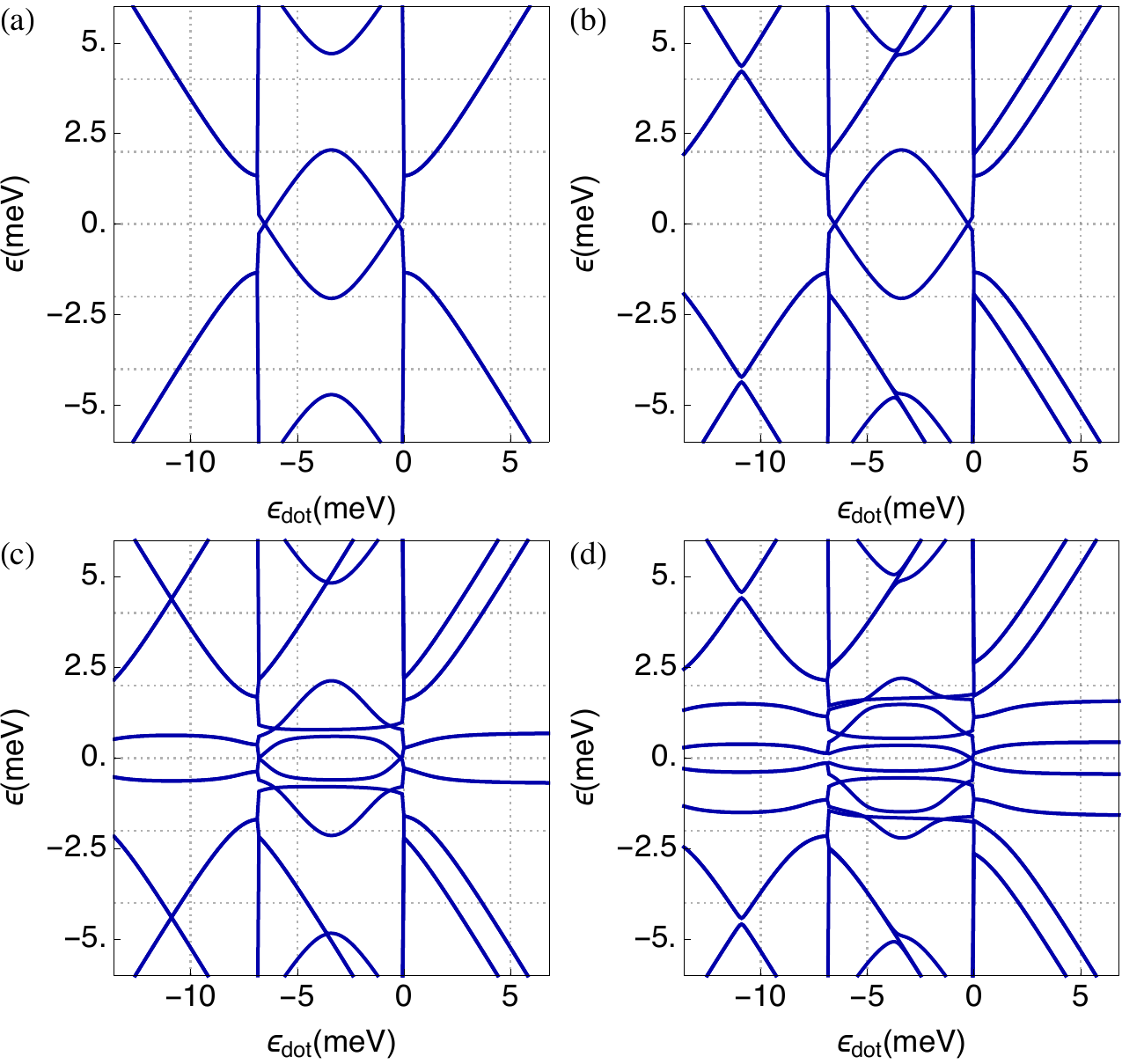}
\caption{Spectra of $H = H_d + H_R + \Sigma_R$ corresponding to the transport simulations in Fig. \ref{fig3}, where the self-energy $\Sigma_R(\omega)$ of Eq. 
(\ref{Sigma}) is taken at $\omega=0$.}
 \label{figsupp8}
\end{figure}

For simplicity, we assume that the ABS are uniformly coupled to the dot states through a spin-independent hopping amplitude
\begin{equation}
V=t\sum_{ij\sigma}\,d^\dagger_{i\sigma} c_{j\sigma} + \mathrm{h.c.},
\end{equation}
This minimal model for the quantum dot coupled to a finite-size superconducting lead fully captures the physics resulting from a microscopic model describing a quantum dot coupled to a finite proximitized nanowire (not shown).

The density of states at the dot is obtained by taking the trace over electron-like dot states $\rho(\omega) =-\frac{1}{\pi}\mathrm{Im}\mathrm{Tr_{dot}}G$ of the retarded Green function $G = [\omega+i0-H-\Sigma_R(\omega+i0)]^{-1}$.

Tunneling spectroscopy is performed through the left superconducting lead, that is assumed to be tunnel-coupled to the dot. Using the Keldysh formalism, it is possible to write an expression for the current in terms of Green's functions of the above model (for details, see e.g. the supplemental information of Ref. [\onlinecite{LeeNatnano2014}]. For very asymmetric situations (such that the one described here where the left superconducting lead just acts as a tunneling probe), the lengthy Keldysh expression can be simplified and one can derive a tunnel-like expression of the current which is just a convolution of the DOS in the probe and the DOS at the quantum dot of the form
\begin{equation}
I\propto \int \rho(\epsilon)\rho_{probe}(\epsilon+eV)[f(\epsilon)-f(\epsilon+eV)]d\epsilon.
\end{equation}
The density of states in the probe contains a smooth gap, modelled as a Dynes-BCS function $\rho_{probe}(\epsilon)=\mathrm{Re}\,\rho_\mathrm{BCS}(\epsilon+i\eta_D)$, with a phenomenological Dynes depairing imaginary part $\eta_D=0.01$ meV and
\begin{equation}
\rho_\mathrm{BCS}(\epsilon)=\rho_0\frac{|\epsilon|}{\sqrt{\epsilon^2-\Delta_p^2}},
\end{equation}
with $\rho_0$ the DOS at the Fermi energy of the probe in the normal state.
In simulations with a zero bias peak in the probe, the latter is added as a Gaussian of width $0.25$ meV.

\begin{figure*}[h!]
  \centering
  \includegraphics[width=1.0\textwidth]{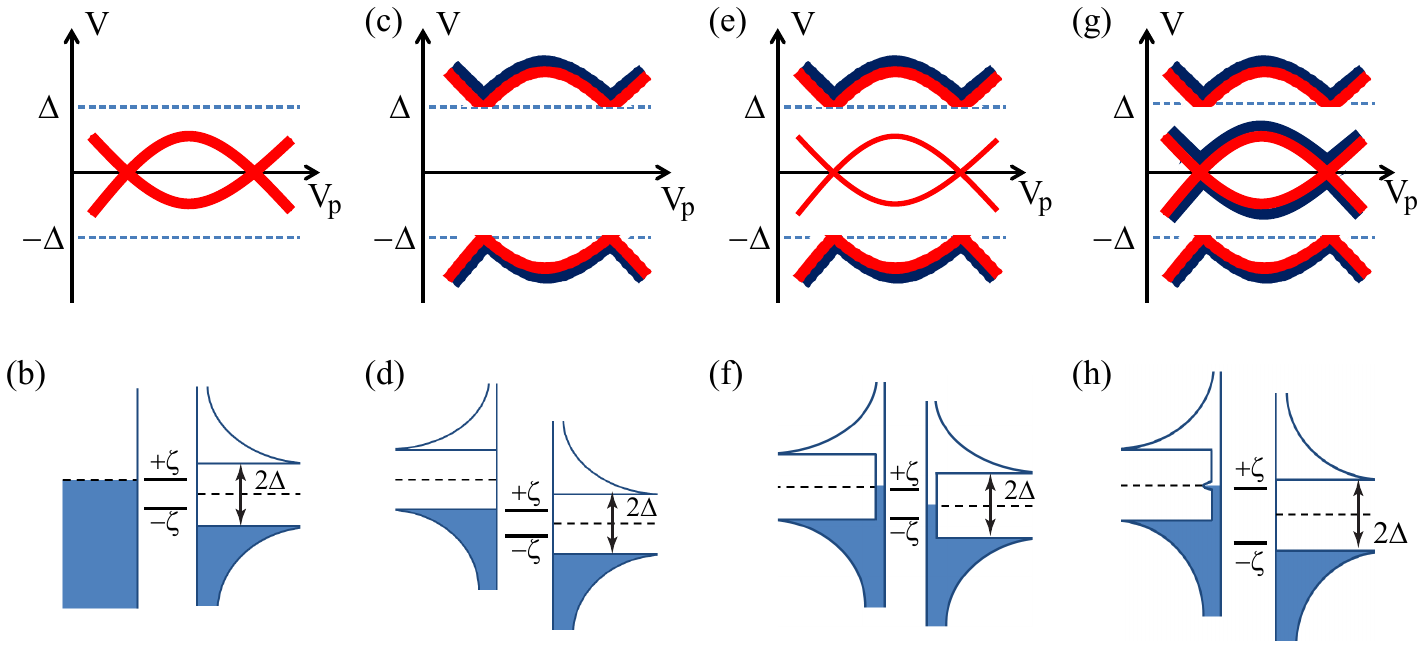}
  \caption[Illustrative bias vs. gate plots with different tunneling probes.]{\textbf{Illustrative bias vs. gate plots with different tunneling probes.} 
  (a) Bias vs. gate plot with a normal conductor probe. Positive dI/dV resonances are depicted in red. (b) Energy diagram with ABS energy $+\zeta$ at resonance with the normal probe. (c-d) Bias vs. gate plot and energy diagram with a hard gap superconducting probe. The blue curves in (c) depict negative dI/dV resonances. (e-f) Bias vs. gate plot and energy diagram with a soft gap superconducting probe (a uniform DOS within the gap). (g) Bias vs. gate plot and energy diagram relevant for experiments on NbTiN/InSb devices. The soft gap probe has a DOS peak at zero bias, which produces NDC at low bias in (g). 
  }
 \label{supp1}
\end{figure*}

\begin{figure*}[h!]
  \centering
  \includegraphics[width=0.7\textwidth]{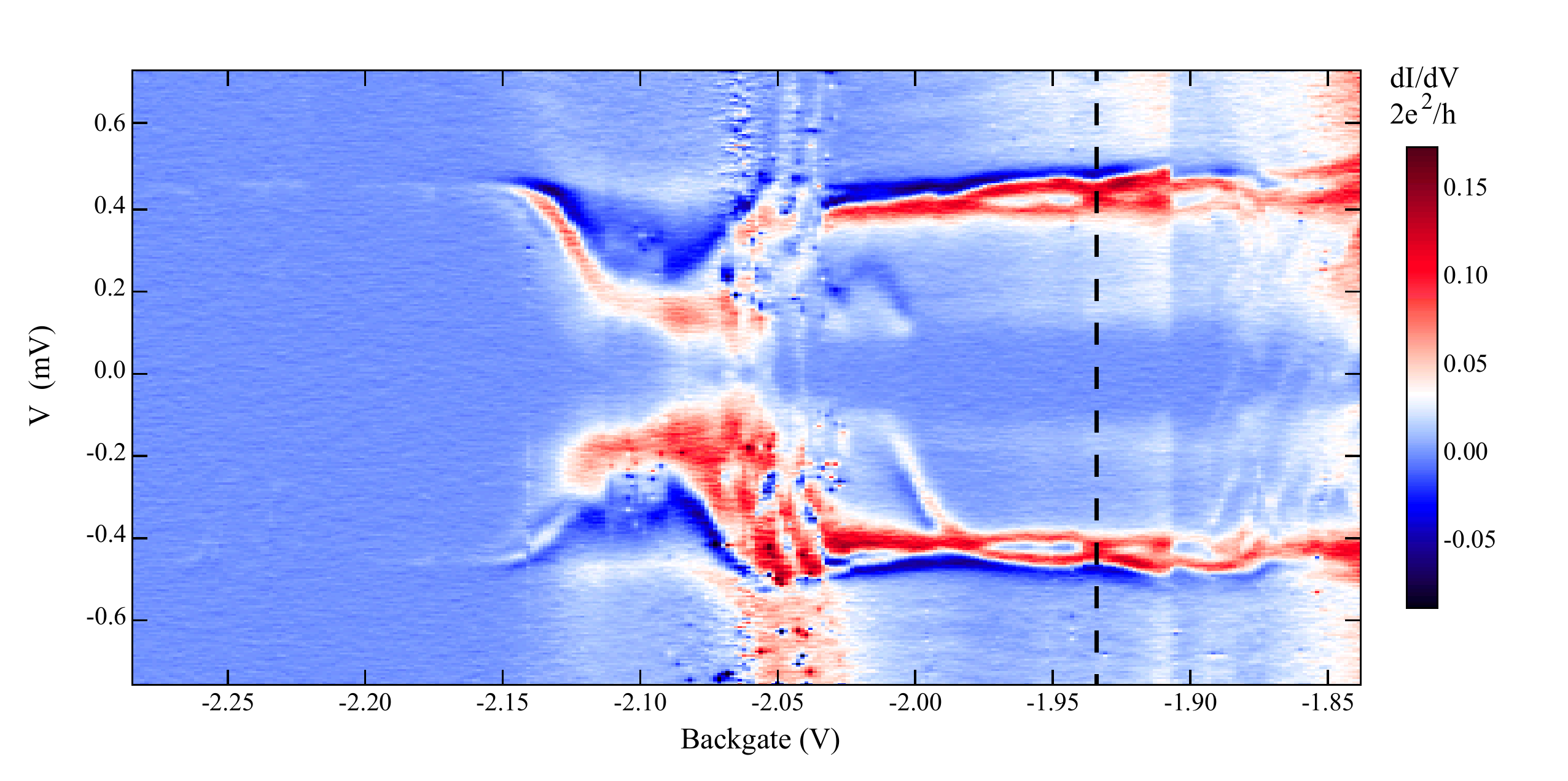}
  \caption[Gate scan device 1]{ 
  \textbf{Back gate scan for Al/InSb device discussed in the main text.} Differential conductance through the device as a function of bias and backgate voltage shows the unintentional quantum dot present in the nanowire. The linecut corresponds to the backgate voltage at which the data in Fig. 1 of main text are taken. Data obtained at zero applied magnetic fields. At negative backgate voltages below -2.15 V, current-voltage characteristics resemble a hard gap regime, however the induced gap likely remains soft and the suppressed conductance within the gap is due to an increased tunnel barrier.
}
 \label{supp2}
\end{figure*}

\begin{figure*}[h!]
  \centering
  \includegraphics[width=0.7\textwidth]{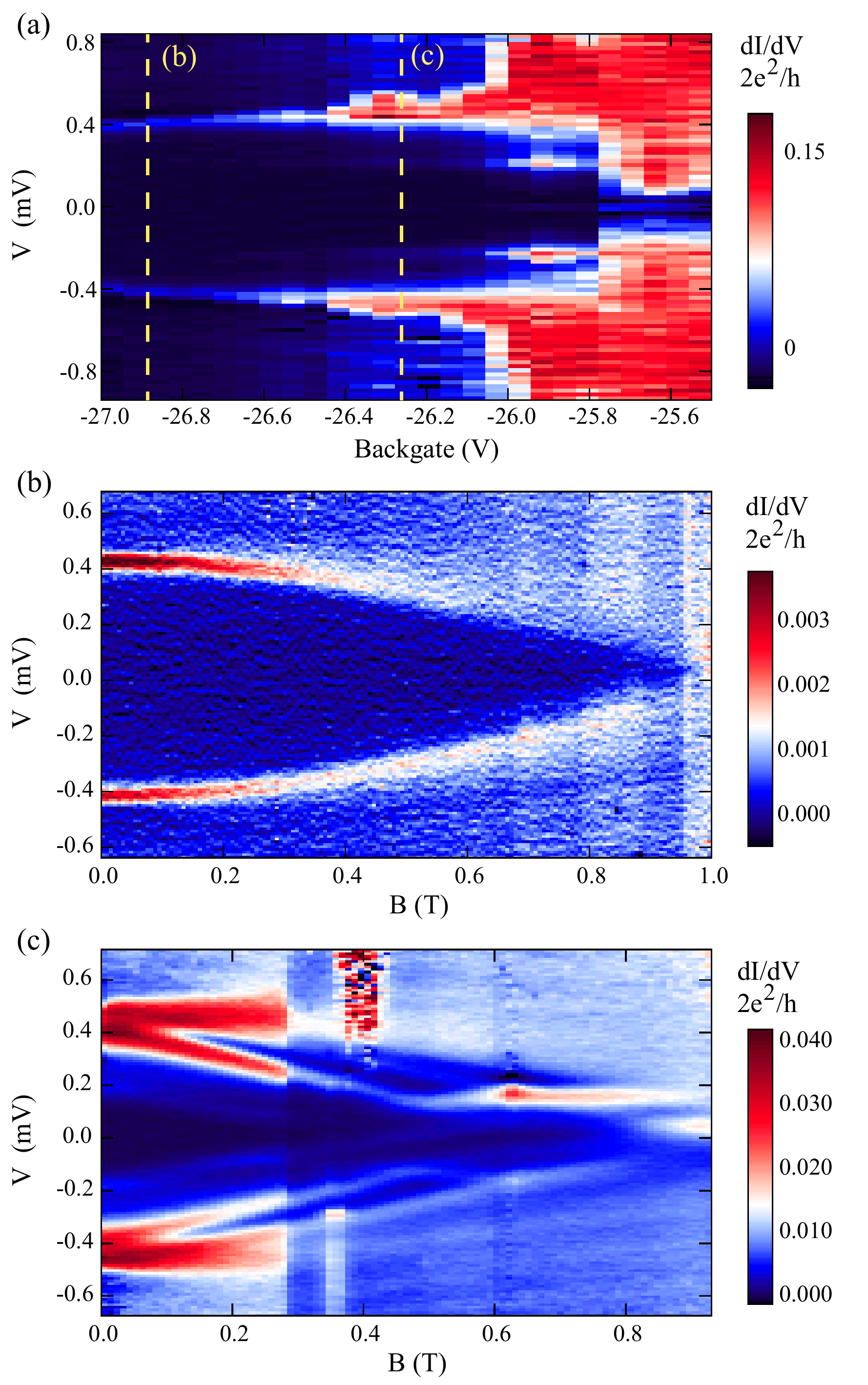}
  \caption[Device 3 (InSb/Al)]{ 
  \textbf{Data from a similar Al/InSb device that is not discussed in the main text.} (a) Tunneling conductance through the device as a function of bias and backgate voltage. Gate voltages corresponding to panels (b) and (c) are marked with vertical dashed lines. (b) Superconducting gap with no apparent additional resonances persists up to $B=1$ T. In agreement with data in Fig.\ref{supp2}, data obtained at more positive gate voltages, with more open tunnel barrier, reveal subgap structure due to soft gap: (c) There are two resonances deviating from the gap edge at finite magnetic field as well as apparent replicas of those resonances at half of the gap energy.}
 \label{supp3}
\end{figure*}

\begin{figure*}[h!]
  \centering
  \includegraphics[width=0.9\textwidth]{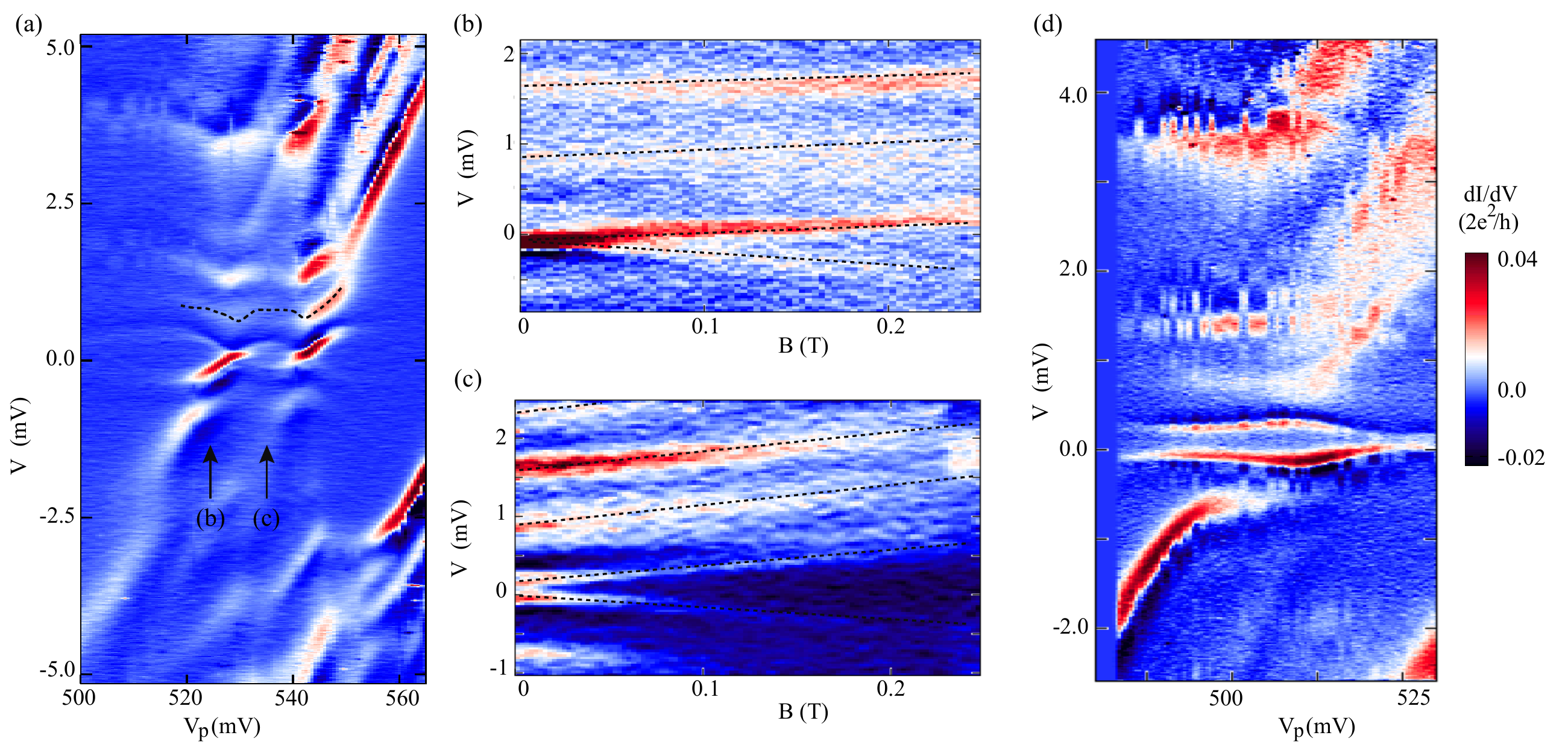}
  \caption[Replicas of subgap resonance at high bias.]{\textbf{Replicas of subgap resonance at high bias.} 
  (a) Data from Figure 2 in the main text. One replica is traced by the dotted line. Gate settings that correspond to panels (b) and (c) are marked by arrows. (b) Magnetic field vs bias dependence at the gate position marked with (b) in panel (a). Note that at this gate position, only the negative bias branch of the lowest resonance has high conductance due to asymmetric dot barriers. (c) Magnetic field vs bias dependence  at the gate position marked with (c) in panel (a). The interesting aspect of magnetic field dependences presented here is that all replicas appear to shift in the same direction for positive bias. Magnetic field evolution of ABS has not been studied in the theoretical model. (d) Bias vs. gate measurement in the open-dot regime showing anticrossing ABS rather than loops. These features are also accompanied by high bias resonances which are qualitatively similar to replicas discussed in the main text.
  }
 \label{supp4}
\end{figure*}

\begin{figure*}[h!]
  \centering
  \includegraphics[width=0.5\textwidth]{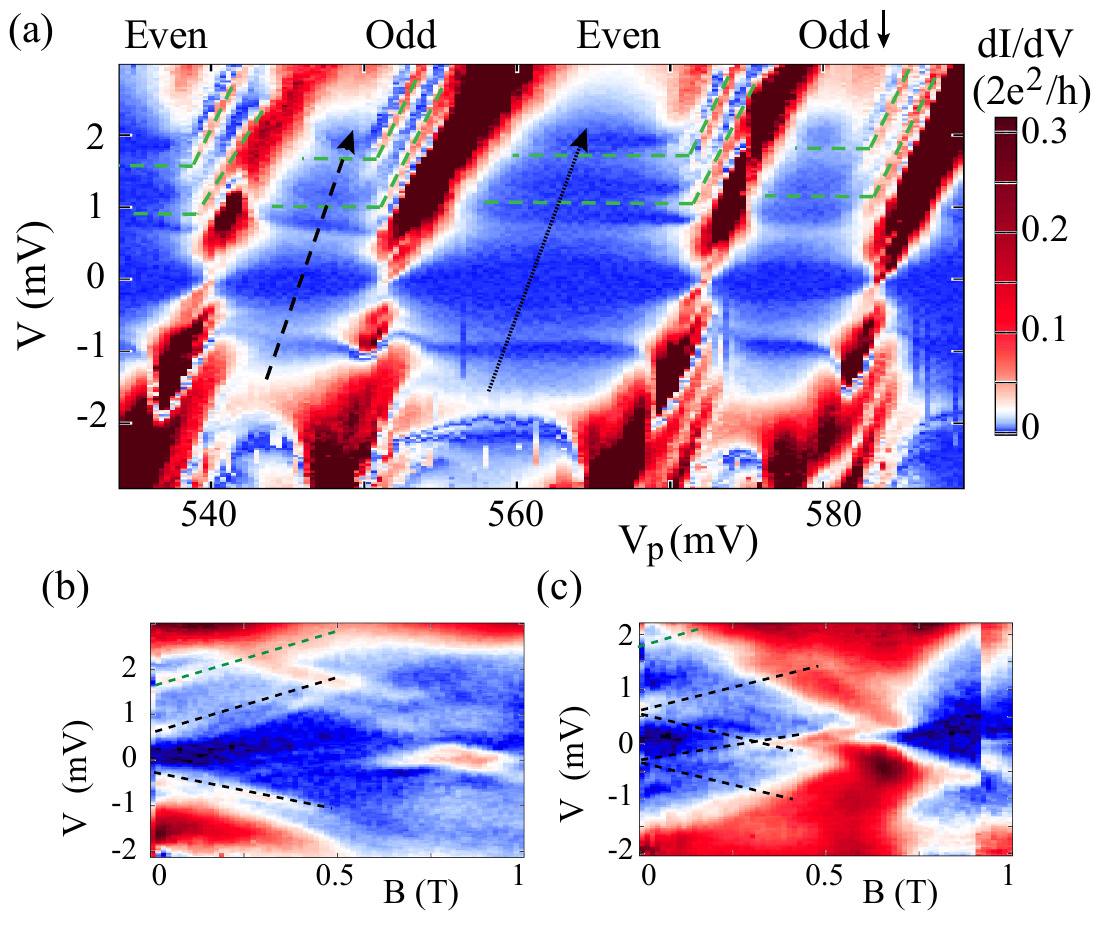}
  \caption[Horizontal resonances.]{\textbf{Horizontal resonances.} 
  (a) Bias vs. gate measurement in the co-tunneling regime with the quantum dot decoupled from both superconductor contacts. $V_s$ = 50 mV and $V_t$ = -500 mV. Horizontal resonances at multiple biases are observed. They appear to be extensions of the diagonal lines, depicted by the green dashed lines. These data confirm that replicas are related to excited states in quantum dots. ``Even'' and ``Odd'' mark tentative parities of dot occupations. The black arrows indicate bias sweeps used in panels (b) and (c). (b) and (c) Magnetic field dependence of the resonance peaks in an odd and even occupation diamond, respectively. Black dashed lines track the field evolution of the lowest resonances while green dashed lines track the field evolution of the high bias resonances. A zero-bias peak is observed in panel (b) over a significant range of magnetic field, however its' connection to MBS is unlikely.
}
 \label{supp7}
\end{figure*}

\begin{figure*}[h!]
  \centering
  \includegraphics[width=0.7\textwidth]{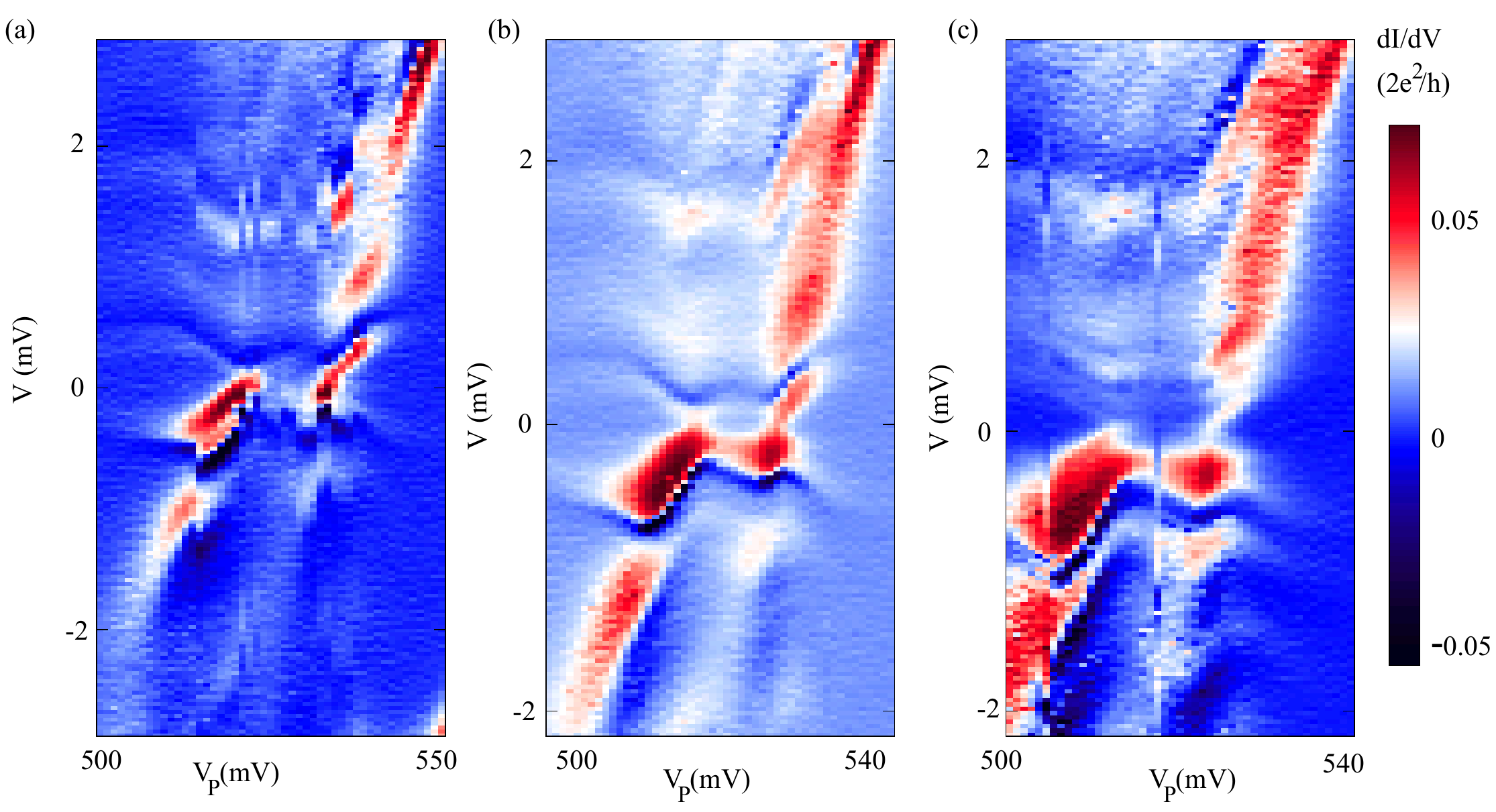}
  \caption[The effect of $V_t$ in the closed dot regime.]{\textbf{The effect of $V_t$ in the closed-dot regime.} 
  $dI/dV$ as a function of bias and $V_p$. $V_s$ = 220 mV. (a) $V_t$ = -685 mV. (b) $V_t$ = -670 mV. *c) $V_t$ = -655 mV. The gate $t$ is expected to tune the DOS in the probe. Over this range of $V_t$, subgap NDC is always present ruling out an intermittent state at zero bias in the tunneling probe. 
}
 \label{supp5}
\end{figure*}

\begin{figure*}[h!]
  \centering
  \includegraphics[width=1.0\textwidth]{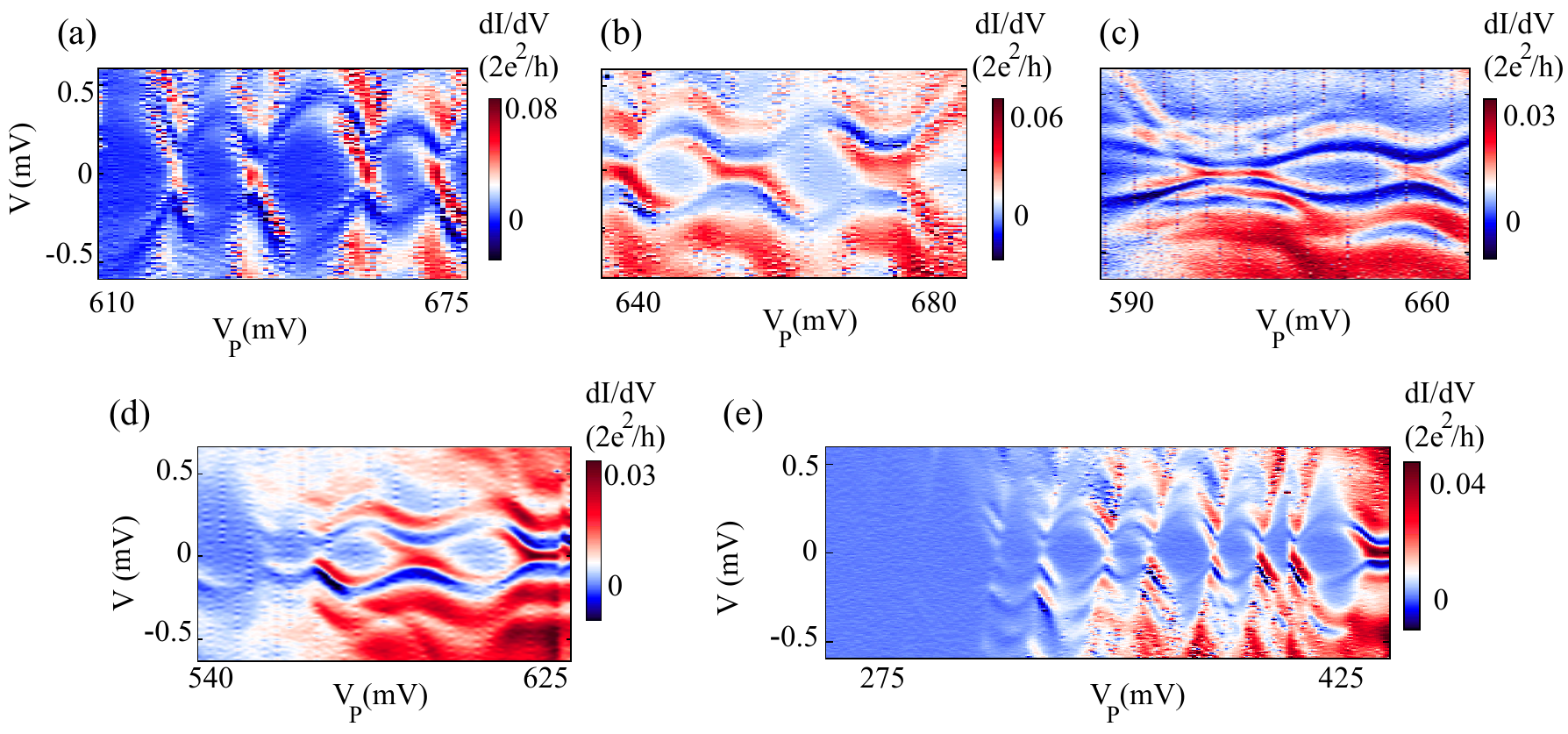}
  \caption[Another dot created with the same device. ]{\textbf{Another dot created with the same NbTiN device. } 
This dot is strongly coupled to the left superconducting lead and tunnel-coupled to the right lead. The data show similar features such as multiple apparent replicas of loop resonances connecting to diagonal resonances, as well as subgap NDC. $dI/dV$ scans as a function of bias and $V_p$ at various $V_s$ and $V_t$. (a) $V_s$ = 50 mV and $V_t$ = -700 mV. (b) $V_s$ = 150 mV and $V_t$ = -700 mV. (c) $V_s$ = 225 mV and $V_t$ = -700 mV. (d) $V_s$ = 300 mV and $V_t$ = -700 mV. (e) $V_s$ = 150 mV and $V_t$ = -600 mV. 
}
 \label{supp6}
\end{figure*}

\end{document}